\documentclass{aa}  

\usepackage{lipsum}
\usepackage{verbatim}
\usepackage{float}
\usepackage{graphicx}
\usepackage{hyperref}
\usepackage{placeins} 
\usepackage{txfonts}

\newcommand*{\ditto}{--- \raisebox{-0.75ex}{\texttt{"}} ---}
\newcommand{\phs}{\mbox{${\rm ph}\,{\rm s}^{-1}{\rm m}^{-2}{\rm \mu m}^{-1}{\rm arcsec}^{-2}$}}
\newcommand{\magarcsec}{\mbox{mag\,arcsec$^{-2}$}}
\newcommand{\mum}{\mbox{$\mathrm{\mu m}$}}
\newcommand{\px}{\mbox{px}}
\newcommand{\wsqm}{\mbox{W\,m$^{-2}$}}
\newcommand{\nr}{N\textsuperscript{\underline{\scriptsize o}}}
\newcommand{\elec}{\mbox{$\mathrm{e^-}$}}
\newcommand{\epx}{$\mathrm{e^-\,px^{-1}}$}

\begin{document} 

\title{The near infrared airglow continuum conundrum}

\subtitle{Constraints for ground-based faint object spectroscopy}

\author{J. K. M. Viuho\inst{1,2,3,4}
\and
J. P. U. Fynbo\inst{1,2}
\and
M. I. Andersen\inst{1,2}
}

\institute{The Cosmic Dawn Center (DAWN)
           \and
           Niels Bohr Institute, University of Copenhagen, Jagtvej 155A, DK-2200, Copenhagen N, Denmark
           \and
           Nordic Optical Telescope, Rambla José Ana Fernández Pérez 7, ES-38711 Breña Baja, Spain
           \and
           Department of Physics and Astronomy, Aarhus University, Munkegade 120, DK-8000 Aarhus C, Denmark\\
           \email{joonas.viuho@nbi.ku.dk}
           }

\date{Received Month dd, YYYY; accepted Month dd, YYYY}

 
\abstract
{The airglow continuum in the near infrared is a challenge to quantify due to its faintness, and the grating scattered light from atmospheric hydroxyl (OH) emission lines. Despite its faintness, the airglow continuum sets the fundamental limits for ground-based spectroscopy of faint targets, and makes the difference between ground and space-based observation in the interline regions between atmospheric emission lines.}
{We aim to quantify the level of airglow continuum radiance in the VIS -- NIR wavelength range observable with silicon photodetectors for the site Observatorio del Roque de los Muchachos in a way that our measurement will not be biased by the grating scattered light. We aim to do this by measuring the airglow continuum radiance with a minimal and controlled contamination from the broad instrumental scattering wings caused by the bright atmospheric OH lines.
}
{We measure the airglow continuum radiance with longslit $\lambda/\Delta\lambda\sim4000$ spectrograph in $\sim$100\,{\AA} wide narrow band passes centered at 6720, 7700, 8700 and 10\,500\,{\AA} (in line with the R, I, and Z broadbands) with the 2.5-meter Nordic Optical Telescope under photometric dark sky conditions. The bandpasses are chosen to be as clean as possible from atmospheric absorption and the OH line emission keeping the radiation reaching the grating surface at minimum.}
{We observe the zenith equivalent airglow continuum to be 22.5\,\magarcsec{} at 6720\,{\AA} band, and 22\,\magarcsec{} at 8700\,{\AA}. We derive upper limits of 22\,\magarcsec{} at 7700\,{\AA} due to difficulty to find a clean part of spectrum for measurement, and 20.8\,\magarcsec{} at 10\,500\,{\AA} due to low system sensitivity. Within measurement errors and the natural variability expected for the airglow emission our results for the Observatorio del Roque de los Muchachos are comparable to values reported for other major observatory sites. With our medium resolution spectra we are unable to comment on the origin of the radiance which can still be due to faint unresolved spectral lines or the true (pseudo)continuum. The measurement uncertainty on the zenith scaled continuum radiance is dominated by the detector effects, assumptions on atmospheric scattering, and choice of zodiacal light model. 
}
{We conclude the airglow continuum radiance not to be due to instrumental effects in our bandpasses, and we measure it to be two to four times brighter than the zodiacal light towards the ecliptic poles, the darkest foreground available for both ground- and space-based observatories. While the level is not negligible, it is dark enough to encourage investigating novel optical technologies and applying already known stray light reduction techniques to the future NIR and SWIR spectroscopic instrumentation.
}

\maketitle
 %

\section{Introduction}
\label{sec:intro}
{\let\thefootnote\relax\footnotetext{We define the spectral ranges referred often in the following way: visible 4000--7000\,{\AA} (VIS), near infrared (NIR) 7000--11\,000\,{\AA}, and short-wave infrared (SWIR) 1.1--3\,\mum{}. All magnitudes in our work refer to magnitudes in the AB system.}}

Astronomy has entered to photon noise limited era with ever lower noise photo-detectors and larger aperture telescopes, highlighted by the forthcoming Extremely Large Telescopes (ELTs). Both ground- and space-based observatories are affected by the Zodiacal Light (ZL), while ground-based facilties see additional emission from the foreground sky, of which continuum radiance is still not well understood. To set a reference for the foreground brightness, the darkest VIS--NIR foreground available from both ground and space is towards the ecliptic poles, making them ideal zones of exploration for deep fields such as the Hubble Deep Field \citep[HDF;][]{Williams96}, and Great Observatories Origins Deep Survey \citep[\mbox{GOODS};][]{Giavalisco04} fields. The yearly average ZL foreground in HDF is $\sim$50\,\phs{} at 1.25\,\mum{} corresponding to 22.4\,\magarcsec{}. Similarly, the Cosmic Evolution Survey \citep[\mbox{COSMOS};][]{Scoville07, Weaver22} field located at a lower ecliptic latitude, \mbox{$\beta_{\rm ec}=-9\degr$}, offers a background of $\sim$ 85\,\phs{} or 21.8\,\magarcsec{} for the darkest part of the year, but undergoes large variation depending on the Solar elongation angle. Ground based astronomical observatories are located at some of the darkest sites around the globe with typical broadband sky brightness ranging in visual (VIS) V $\sim$ 22\,\magarcsec{}, but rapidly increasing towards longer wavelengths, reaching I $\sim$ 20\,\magarcsec{} in near infrared (NIR), and J $\sim$ 16\,\magarcsec{}, H $\sim$ 14\,\magarcsec{} in short-wave infrared (SWIR). The reason for the increase is the radiance from increasing number density, and brightness of rotation-vibrational transitions of hydroxyl (\mbox{OH}) molecules, and to lesser extent molecular oxygen \mbox{O$_2$}. The behavior of atmospheric line emission is understood to a level where sophisticated models have been developed \citep[e.g. ESO SkyCalc,][]{Noll12,Jones13}. However, the interline airglow continuum emission, especially in NIR--SWIR wavelengths, remains much less well known due to its faintness and challenges in its measurement. Instrumental effects such as grating scattered light and thermal stray light complicate the airglow continuum measurement if the experiment measuring it is not designed in detail.

Published dedicated airglow continuum radiance measurements at VIS and NIR wavelengths are mainly from the 1960's and 1970's \citep{Krassovsky62, Broadfoot68, Sternberg72, Gadsden73, Noxon78, Sobolev78}. Since then, the VIS and NIR continuum radiance has been studied at Cerro Paranal, Chile \citep{Hanuschik03,Patat08,Noll24}. After introduction of SWIR Mercury-Cadmium-Telluride -detectors, the interest in airglow continuum radiance shifted mostly to SWIR \citep{Maihara93b, Cuby00, Ellis12, Sullivan12, Trinh13, Oliva15, Nguyen16} where the ZL foreground would be significantly lower than in NIR \citep{Leinert98, Windhorst22}. Special attention has been given to a relatively narrow spectral region in H-band located around 16\,650\,{\AA}, originally selected by \cite{Maihara93b}, despite later studies finding few emission lines in this region \citep{Oliva15}. We summarize previous airglow continuum measurements found in the literature in Table \ref{table:previous}.

Both spectroscopic and narrow band imaging measurements have been done to determine the airglow continuum radiance in NIR and SWIR range. Both of the methods having their advantages and disadvantages: it is difficult to find a clean spectral bandpass without airglow emission lines for narrow band imaging, and grating spectrographs are susceptible to suffer from grating scattered light, which can appear as dislocated copies of a parent line and broad diffuse line wings \citep{Woods94,Koch21}. In NIR and SWIR wavelengths, the scattered light combined from all airglow emission lines can result in an artificial continuum \citep{Ellis08, Sullivan12, Oliva15}. \cite{Sullivan12} concluded that their continuum measurement in the H-band could be explained by the spectrograph's Line Spread Function (LSF) wings, while shorter wavelength bands remain unaffected. Thermal blackbody radiation from the instrument may additionally affect the measurement \citep{Ellis20}. To minimize the effect of grating scattered light, a number of the spectroscopic measurements measurements in the 1--2\,\mum{} range have been accompanied with \mbox{OH} line emission suppression units \citep[e.g][]{Maihara93a, Ellis12}.

A further complication is that the airglow exhibits significant temporal variation with timescales ranging from minutes (mesospheric buoyancy waves) \citep{Smith06,Moreels08}, to years (Solar cycle) \citep{Leinert95, Mattila96,Krisciunas97, Patat08, Noll17}. Additionally, semi-annual \citep{Grygalasvhyly21}, and diurnal variability \citep{Smith12} of line emission is present. The airglow continuum has been observed to exhibit diurnal variability \citep{Trinh13}. The NIR and SWIR \mbox{OH} and \mbox{O$_2$} line emission, are known to originate from the mesopause region \citep[e.g.][]{Baker88}. There is a strong density change in the mesopause, and waves can travel at the boundary layer. Typically, these gravity- or buoyancy-waves can be observed with a $\sim$15\,\% brightness variation on time scales from a few up to tens minutes \citep{Moreels08}. However, mesospheric bore events can drive solitons which may cause localized, sudden large amplitude changes \citep[e.g.][]{Smith06}.

Despite the challenges in measurement, a number of chemo-luminescent processes are known to contribute to the atmospheric continuum radiance. Several nitrogen monoxide (\mbox{NO}) reactions are known to contribute to the total VIS -- NIR continuum \citep[see][and references therein for extended discussion]{Bates93, Khomich08, Noll24}. However, none of the reactions have a rate that is sufficiently high to explain the total continuum radiance that has been observed. Recently, it has been shown that iron monoxide (\mbox{FeO}) has an extended pseudo-continuum emission spectrum extending from VIS to SWIR, with the potential to explain a significant fraction of the VIS--NIR continuum emission \citep{Noll24}. \mbox{FeO} still leaving a significant SWIR component unattributed which is possibly due to hydroperoxyl \mbox{HO$\mathrm{_2}$} \citep[preprint,][]{Noll25}.

The objective of this study is to measure the airglow continuum with a method that strongly reduces the effect of grating scattered light. Our study is limited to the spectral range observable with a CCD, i.e. below about 11\,000\,{\AA}. The airglow spectra were observed through narrow band (NB) filters, which is not a full fledged OH suppression scheme, but allows selection of band pass such that the number count and brightness of unwanted OH line emission stays at minimum. Data recording is described in detail in Sec. \ref{sec:observations}. Due to the faint signal level, additional effort was spent on reducing the data which is discussed in Sec. \ref{sec:reduc}. We try to compensate for the low of sensitivity close to the detector sensitivity limit by observing thicker air columns, i.e. fields at very large zenith distances, and our assumptions that went into the final results are commented on the analysis Sec. \ref{sec:analysis}. Finally, implications of our findings are discussed in Sec. \ref{sec:discussion} and \ref{sec:conclusion}. Observed apparent airglow spectra are included in Appendix \ref{app:apparent}. References to previous similar studies found in literature are gathered in Appendix \ref{app:history} and their results are converted to units used in this work to allow comparison.

\begin{figure*}[!ht]
    \begin{center}
    \includegraphics[width=.8\textwidth]{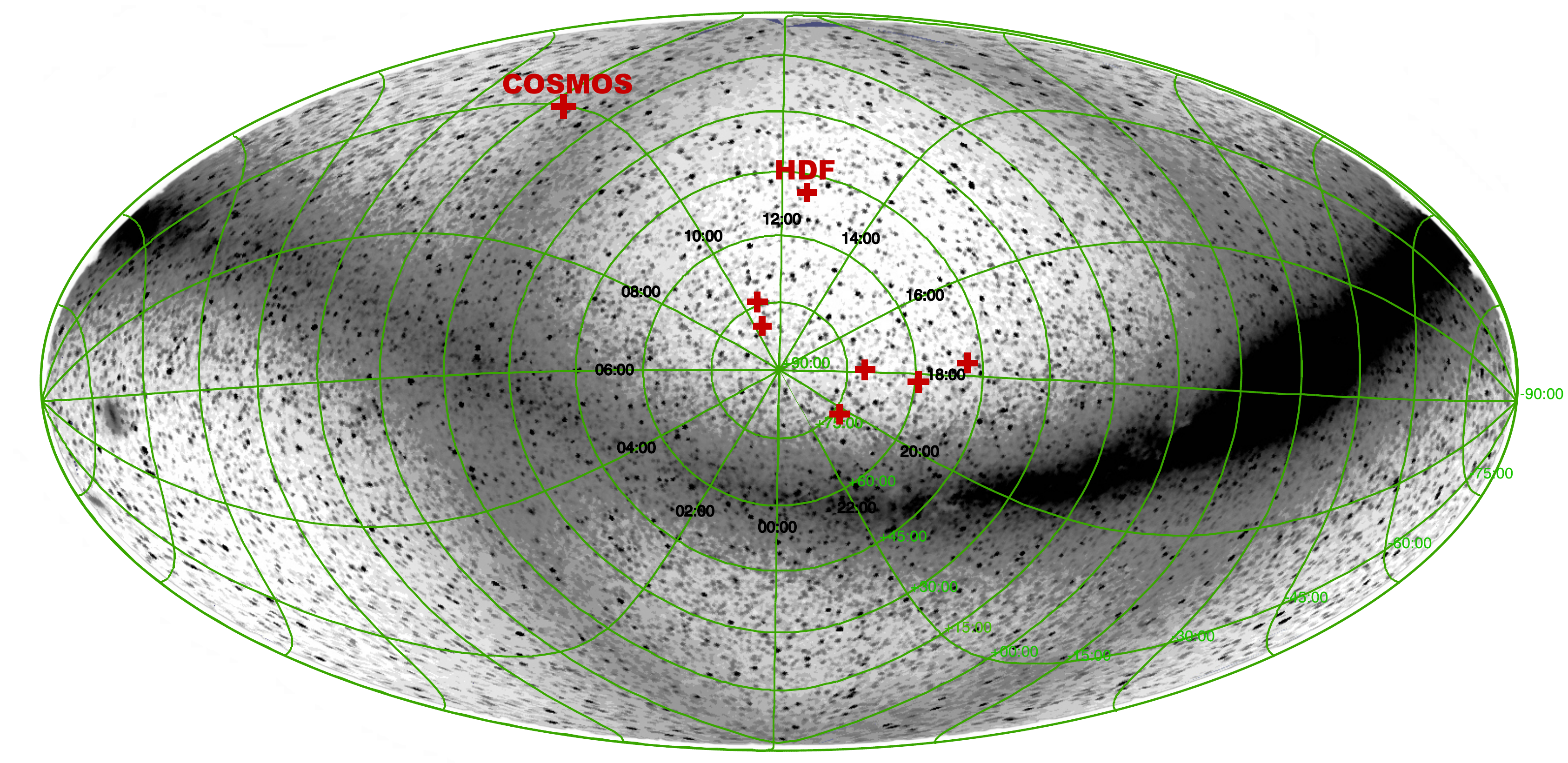}
    \end{center}
    \caption{\label{fig:diffuse}Projection of the COBE/DIRBE Band~1 map with central wavelength at 1.25\,\mum{} towards the ecliptic north pole. The red markers indicate the coordinates for our pointings, the Hubble Deep Field (HDF), and the Cosmic Evolution Survey (COSMOS) field. Milky way crosses the map horizontally while zodiacal light in the ecliptic plane shows as a ring between declinations $\delta=[-30, 30]$.
    }
\end{figure*}

\section{\label{sec:observations}Observations and Data}

\subsection{\label{sec:site}Site}
We recorded airglow continuum spectra with the ALFOSC instrument mounted on the 2.56-meter Nordic Optical Telescope (NOT). The NOT is located at the Observatorio del Roque de los Muchachos on La Palma, Canary Islands, Spain with geographical coordinates {+28\,\degr 45\arcmin 26.2\arcsec\,N}, {17\degr 53\arcmin 06.3\arcsec\,W}, and altitude 2382\,meters. The NOT has a lower pointing limit of {6\degr} above the horizon, and its location is such that the horizon is unobstructed when observing towards the north. The northern direction has the least amount of light pollution on site since the line-of-sight travels fully above the Atlantic, with the only sources of artificial illumination being the town of Roque del Faro 5~km north of the Observatorio del Roque de los Muchachos. Light pollution which is mostly caused by low pressure sodium lamps with characteristic wavelengths of 5890 and 5896\,{\AA} falling outside the band passes in our study. The next source of artificial illumination on the line of sight is the island of Madeira which is 450~km north of La Palma.

\subsection{\label{sec:fields}Field selection}
The observed fields on the sky were chosen to be as dark as possible outside the atmosphere, away from both ecliptic and galactic planes, and yet close to the horizon to be able to observe it with a thick air column. COBE Diffuse Infrared Background Experiment (DIRBE) \citep{Hauser1998, Arendt98, Kelsall98,Dwek98} Band 1 -maps, with a central wavelength of 1.25\,\mum{} were first used to locate fields with low surface brightness, and then PanSTARRS DR1 z-band images \mbox{\citep{Flewelling20}} were used to fine tune the location to the least amount of visible sources and assist orienting the slit on sky. To summarize, the coordinates for the pointings were chosen with the following selection criteria:

\begin{itemize}
    \item Ecliptic latitude $\beta_{\rm ec}>50\degr$.
    \item DIRBE band 1 radiance $<85$\,\phs{}.
    \item No bright, $m$ < 12~mag, stars in the 6.5\arcmin$\times$6.5\arcmin ALFOSC field-of-view.
    \item No source visible in the slit area in PanSTARRS z-band image.
    \item Observable with zenith distance $z>75\degr$ towards the north.
    \item Only a few degrees variation in zenith distance during the measurement.
\end{itemize}

\noindent
Additionally, the following time restrictions were adopted:

\begin{itemize}
    \item Moon altitude < $-$30\degr
    \item Moon separation > 60\degr,
\end{itemize}

\noindent
to make sure that we would not observe scattered Moon light \citep{Krisciunas91,Jones13}. Typically, the Moon illumination fraction during our observations was <2~\%. Additionally, the first observing run purposefully coincided with the Winter Solstice to provide the longest possible night. The time of observations was typically several hours after the sunset, and in most cases past the local midnight.

\subsection{Optical setup}
\begin{figure*}[t]
    \begin{center}
    \resizebox{\hsize}{!}{\includegraphics[width=\textwidth]{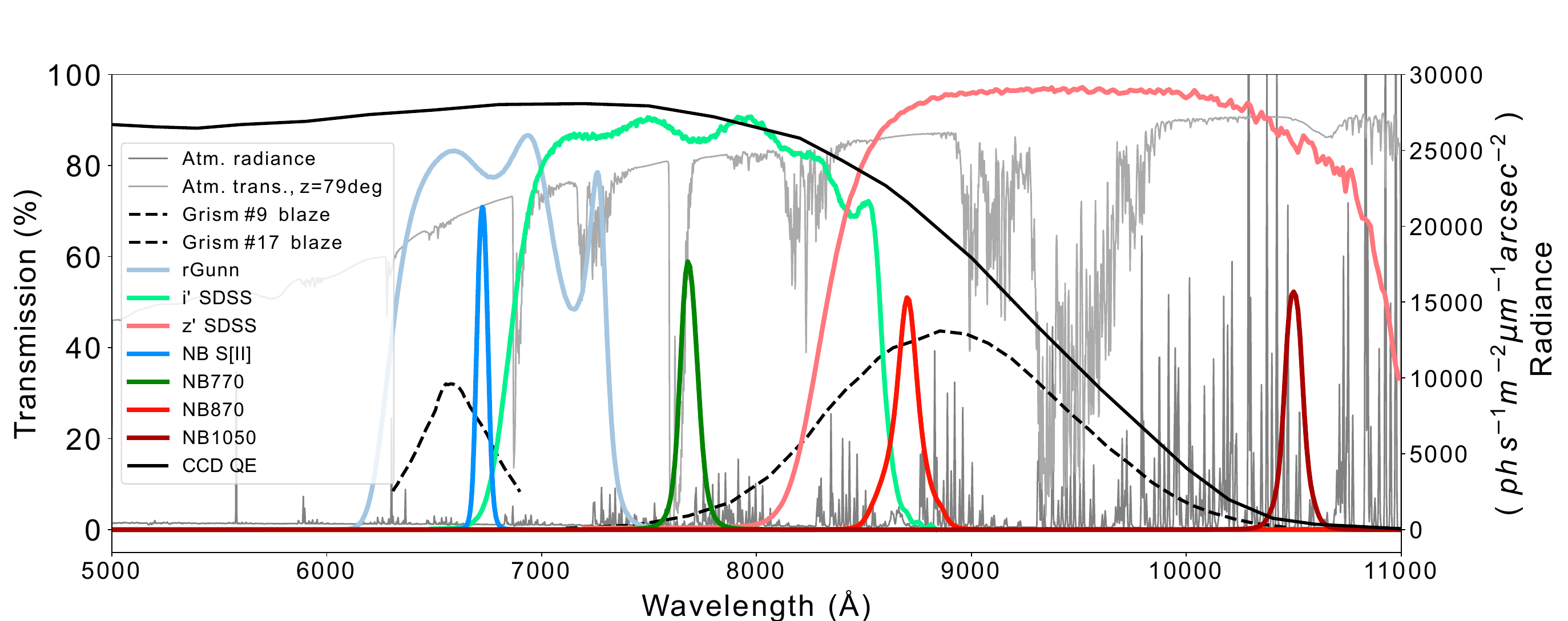}}
    \end{center}
    \caption{\label{fig:filters}Transmission curves of the NB filters, order sorting broad band filters, grism blaze functions, detector efficiency, and the atmospheric transmission at the observed zenith distance. The blue half of the 7700\,{\AA} band is strongly absorbed by atmospheric \mbox{O$_2$}. In the 10\,500\,{\AA} band the total system efficiency is very low since the wavelength falls in the wing of the grism blaze function. 
    }
\end{figure*}

The optical setups that were used are summarized in Table \ref{table:optics} referring to the NOT's optical element database optical component names. Four NB filters were purchased for the experiment, with spectral band passes chosen such that the band passes would be mostly clean from bright (\mbox{OH}) lines, and atmospheric molecular absorption. The chosen bands were centered at 6720, 7700, 8700 and 10\,500\,{\AA} (see Fig. \ref{fig:filters}). The band passes were generally clean from strong absorption apart from the 7700\,{\AA} band which has significant {O$_2$} absorption at the blue side of the band pass. The bands passes were 60--120\,{\AA} in width limiting the total {OH} line flux reaching the grism to its minimum. All bright OH lines within the band passes are located in the wings of the NB filter transmission profiles. NB filters sit in the beam before the grisms on the optical path, thus the stray light introduced by the OH lines should have been kept at its minimum.

The three redder narrow band filters had a diameter of 25.4\,{mm}, which would significantly vignette the ALFOSC beam if placed in standard filter mounting slots, and were consequently mounted on the slit wheel instead. This reduced the effective slit length down to 2.4\arcmin, allowing a sampling of 675\,\px{} on the detector with the reduced slit length. Vertical slits for 6720\,{\AA} band were 5.3\arcmin in length, allowing sampling of 1500\,\px{}. In the three reddest bands, the Echelle Grism \#9 was used as the dispersive element since it had the highest resolution of the available grisms, $\lambda/\Delta\lambda\approx 4300$, with a spectral coverage up to the CCD sensitivity cutoff at 11\,000\,{\AA}. The broadband SDSS i$^\prime$ and z$^\prime$ -filters were used for order sorting with Grism \#9. The 6720\,{\AA} band was observed with Grism \#17 using the Gunn r-band filter for order sorting. The use of the order sorter was probably unnecessary with Grism \#17, but was done as a precaution.

The 2D spectrum was binned along the slit dimension before the readout and, later on during the analysis, the entire remaining slit length was median collapsed into 1D. This process has the potential to cause smearing and additional line broadening. To counter that, both the slits and the grisms were carefully aligned before observation such that the slit and dispersion direction were matching the detector pixel row and column axis to minimize smearing. Both were aligned to less than one pixel RMS difference respective to the detector axes. Potential spread should be negligible, and the possible resulting broadening should be still captured by the line spread function model.

We checked the instrument pixel scale on a few denser standard star fields and found it to be constant over the field-of-view with a value of $0.21377\pm5\times 10^{-5}$~\mbox{arcsec\,px$^{-1}$} by plate solving. Slit widths were measured by illuminating the slit with a standard calibration arc lamp and imaging them with the ALFOSC camera.

\begin{table*}
    \caption{The used optical configurations for narrow band observations.}
    \label{table:optics}
    \centering
    \begin{tabular}{rcccrcclc}
    \hline\hline
                &                        &                  &                 &                    &           &                          &               \\
    Band        & $\lambda/\Delta\lambda$&Slit width        & Slit length     & NB filter name     & FWHM      & Order sorting filter     & Grism         \\
                &                        &                  &                 &                    &           &                          &               \\
    \hline
                &                        &                  &                 &                    &           &                          &               \\
    6720 {\AA}  & 9400                   & 0.51\arcsec      & 5.3\arcmin      & S{[}II{]} 672\_5   &  60\,{\AA} & r Gunn 680\_102         & Grism \#17    \\
    6720 {\AA}  & 3500                   & 1.29\arcsec      & 5.3\arcmin      & S{[}II{]} 672\_5   &  60\,{\AA} & r Gunn 680\_10   2      & Grism \#17    \\
    7700 {\AA}  & 4300                   & 0.45\arcsec      & 2.4\arcmin      & FB770-10           &  80\,{\AA} & i$^\prime$ SDSS 771\_171& Grism \#9     \\
    8700 {\AA}  & 4300                   & 0.45\arcsec      & 2.4\arcmin      & FB870-10           & 120\,{\AA} & z$^\prime$ SDSS 832\_LP & Grism \#9     \\
    10\,500{\AA}& 4300                   & 0.45\arcsec      & 2.4\arcmin      & FB1050-10          & 100\,{\AA} & z$^\prime$ SDSS 832\_LP & Grism \#9     \\
                &                        &                  &                 &                    &           &                          &               \\   
    \hline
    \end{tabular}
    \tablefoot
    {\tiny Component names refer to Nordic Optical Telescope's optical component database names. The pixel scale is $\sim0.2138$~\mbox{arcsec\,px$^{-1}$}.
    }
\end{table*}

\subsection{\label{sec:lsf}Line spread function}
Diffraction theory predicts a Lorentzian-like envelope for the grating distribution function, which will be modulated by the spectrograph camera point spread function. For a good camera optic, the point spread function is approximately Gaussian. In addition to the Lorentzian and Gaussian components, the micro-roughness of the grating surface contributes a diffuse component \citep{Woods94, Koch21}. In order to estimate the wing contribution on the measured continua, a crude estimate of the line spread function (LSF) was made based on Thorium-Argon arc lamp exposures, since no monochromatic light source working within the band passes was available. The approach was far from optimal, due to non-resolved and too faint to fit lamp lines, but was sufficient to derive an upper limit on the grating scattered light. A combination of Lorentzian and Gaussian was fitted on prominent arc lines visible in the band passes. The Grism \#17 was best fitted with a function that was almost completely Lorentzian, whereas the Grism \#9 appeared mainly Gaussian, which might indicate that the acquired Grism \#9 arc exposures were actually not deep enough to expose the Lorentzian wings. However, this would lead fitting the wings to the readout noise floor, thus leading to further overestimation of the scattering wing contribution.

Based on the fits, 95\% of the line flux is within $\pm$0.5\,{\AA}, $\pm$1.7\,{\AA}, and $\pm$2.1\,{\AA} from the line center in the 6720, 7700 and 8700\,{\AA} bands respectively. Similarly, 99\% of the flux is within $\pm$0.9\,{\AA}, $\pm$6.7\,{\AA}, and $\pm$6.6\,{\AA}. To give a gross overestimate of the LSF wing contribution, the remaining 1\% of the light further away from the core can be taken as the diffuse LSF component. Dividing the 1\% of total line flux in each band, and distributing it equally within the FWHM of the NB filter, one reaches an upper limit of 4$\times 10^{-4}$, 0.2, and 0.1\,\epx{} in the 6720, 7700 and 8700\,{\AA} bands respectively, which are well below the systematic uncertainty of 0.6\,\epx{} from bias subtraction (see Section \ref{sec:reduc}). Consequently, we expect the OH line wing contribution originating from the OH lines within the band pass to the measured continuum to be negligible. Our measurement in 10\,500\,{\AA} band is detector noise limited and no continuum detection was reached. Consequently, we did not study the effect of the LSF wings in this band.

\subsection{Detector setup}
ALFOSC is equipped with a deep depleted e2v CCD231-42 backside illuminated CCD with active pixel dimensions of $2048\times2064$\,\px{}. Outside the image area, there is an additional 38\,\px{} of vertical and 2$\times$50\,\px{} of horizontal overscan. The CCD controller has four readout amplifiers, with amplifiers B and D showing very similar performance. For an unknown reason, the binning for amplifier D was limiting the usable detector area, leading us to use amplifier B for the measurement. Also, amplifier B was located further away from the region of interest for the three reddest bands giving the controller more time to stabilize during a readout. Based on NOT's long-term quality control monitoring, the amplifier B readout noise was $4.40\pm3\times 10^{-2}$\,\elec{} for the slowest available read-out speed of 100\,\mbox{kHz}. According to the same quality control the gain was $0.169\pm 2\times 10^{-3}$\,\mbox{\elec{}\,ADU$^{-1}$}. These values are adopted in our analysis. Different binning factors were studied before beginning the observations and a binning of 5 was adopted in the slit dimension.

\subsection{Atmospheric conditions}
All observations were carried out under photometric conditions. The appearance of high clouds along the line of sight were monitored with all sky cameras to make sure that a cloud layer would not interrupt our observations. Low dust particle counts, e.g. Calima from the Sahara \citep{Murdin86}, were measured at the level of the observatory during all of the nights. Tenerife $\sim$130\,{km} away was clearly visible on all nights of observations, giving a visual reference on atmospheric turbidity indicating good atmospheric transparency.

The first set of observations were carried out on the nights of 30 Dec 2021 through to 2 Jan 2022, and a second set of observations were performed in March 2023 (see the list of observations in Table \ref{table:obs}). The main run took place one week after the Winter Solstice during dark sky conditions with no Moon illumination. Observations were made between 00:00\,UT and 02:00\,UT with an altitude of 10$\degr$ above the horizon to guarantee a thick air column and only a small amount of non-atmospheric emission. The horizon was clear during all four December nights and observing conditions were excellent in general. Additional observations were carried out in September 2024, again under clear sky conditions with no Moon illumination. The additional observations were carried out at smaller zenith distances to avoid the Milky Way.

\subsection{Solar conditions}
Data on solar and space weather conditions are presented in Table \ref{table:solardata}, and shown in scatter plot Fig. \ref{fig:solarcorr}. The solar 10.7\,\mbox{cm} radio flux, or $F_{10.7}$ \citep{Tapping13}, data is from the automated solar radio flux monitors of the Dominion Radio Astrophysical Observatory (DRAO) in Penticton, Canada. DRAO reports $F_{10.7}$ three times per day, and the Table \ref{table:solardata} refers to DRAO \emph{observed} flux recorded closest to the time of our observation. The last observation of the day is recorded either at 22:00\,UT (Nov -- Feb), or 23:00\,UT (Mar -- Oct). Solar X-ray data is from the Geostationary Operational Environmental Satellite (GOES), and the solar wind data from the Deep Space Climate Observatory (DSCOVR). Solar and space weather communities typically refer X-ray fluxes in units of X-ray flare class. Flare classes are A, B, C, M and X, referring to $10^{-8}$, $10^{-7}$, $10^{-6}$, $10^{-5}$, and $10^{-4}$\,\wsqm{} in wavelength range of 1--8\,{\AA} respectively. For example, B2.5 indicates $2.5\times10^{-7}$\,\wsqm{}. We will use these units in this work to report solar X-ray background, and solar flares strengths.

From 31 Dec 2021 to 3 Jan 2022, solar activity was generally low with a X-ray background level ranging from B2.5 in the beginning of the run, to B1.5 at the end. On 31 Dec five C-class flares took place, and on 1 Jan one C-class and one M-class flare. $F_{10.7}$ decreased from about 103 to 90\,sfu. The solar wind speed and density varied during the run both were generally increasing towards the end of the run. On 2 Jan 2023, just after UT midnight solar wind speed and density showed a sudden increase, coinciding with the highest measured continuum level at 8700\,{\AA} (obs. ID\,6, see Fig. \ref{fig:app3}), and elevated OH line emission in 6720\,{\AA} band (obs. ID\,5, see Fig. \ref{fig:app1}). In March 2023, the $F_{10.7}$ flux was at a similar level of 90\,sfu as during the first run. However, the X-ray background was an order of magnitude higher averaging at C1. Day of 20 Mar 2023 saw three C-class and one M-class flare. Five C-class flares were recorded on the following day, 21 Mar 2023. The solar wind was stable with comparable values to the first run. The final September 2024 run coincided with higher solar activity with $F_{10.7}$ staying around 230\,sfu, and strong X-ray background ranging between C3 and C4. 1 Jan 2024 saw seven C-class and four M-class flares, strongest being M5.57, and lasting for an extended period of time. The solar wind speed and density were low compared to the the earlier runs.

\subsection{Stellar contamination}
Due to the large zenith distances of the observations and very barren fields, it was difficult to find guide stars to correct the telescope tracking. Most of the observations relied on telescope blind tracking apart from the night of 1 Sep 2024 when telescope was guiding on all pointings. The NOT blind tracking accuracy is reported to be 0.17\,{arcsec\,min$^{-1}$} at $z=14$\degr, and 0.6\,{arcsec\,min$^{-1}$} at $z=70$\degr. The blind tracking error has not been measured at the zenith distances of our observations, but it can be assumed to be larger than the measured error at $z=70$\degr. Assuming a drift of 1\,{arcsec\,min$^{-1}$} at $z=80$\degr, it would take 30\,{s} for a star to cross the slit. This corresponds to less <5\% of the total exposure time. In the data reduction, the entire slit length is median collapsed into 1D. In our analysis we assume that stars potentially crossing the slit due to tracking errors have not contaminated the measured sky radiance.

\subsection{Standard stars}
\label{sec:standards}
One flux standard star observation was taken for each filter once per run. Standard stars were observed at low a zenith distance with a 10\arcsec -slit, or on some occasions without a slit, to determine the system sensitivity function. During the first run in New Year's 2021 - 22, white dwarf \object{LAWD~23} was observed in the 8700\,{\AA} band, and \object{HD~84937} in the rest. During the 2023 run \object{HD~93521} was observed in the 7700, 8700 and 10\,500\,{\AA} bands, and on 1 Sep 2024 \object{HD~19445} was observed in 6720 and 8700\.{\AA} bands. LAWD~23 reference fluxes were taken from \cite{Oke74} corrected by a zero point offset of 0.04~mag as suggested by \cite{Colina94}. Reference fluxes for HD~84937 and HD~93521 were taken from \cite{Rubin22}. No references fluxes were found for any of the three observed standard stars at 10\,500\,{\AA}. Consequently, a spectral template of a O9\,V star \citep{Pickles98} was taken and its Wien's tail beyond 9700\,{\AA} was scaled to match the \cite{Rubin22} HD~93521 flux. HD~93521 has spectral type O9\,III, and it is somewhat cooler than the spectral template used. The flux calibration in the 10\,500\,{\AA} band should only be taken as indicative. Reference flux for HD~19445 was obtained from X-shooter Spectral Library Data Release 3 \citep{Verro22}.

\begin{table*}
    \caption{\label{table:obs}Table of observations.}
    \centering
    \resizebox{\textwidth}{!}{%
    \begin{tabular}{cccrcccccrccc}
        \hline\hline
        \multicolumn{1}{l}{}& \multicolumn{1}{l}{} & \multicolumn{1}{l}{} & \multicolumn{1}{l}{}     & \multicolumn{1}{l}{} & \multicolumn{1}{l}{} & \multicolumn{1}{l}{} & \multicolumn{1}{l}{} & \multicolumn{1}{l}{} & \multicolumn{1}{l}{}         & \multicolumn{1}{l}{} &          \\
        ID                  & Band                 & Time                 & \multicolumn{1}{c}{Date} & Start                & $t_{\rm exp}$        & Time after           & RA                   & DEC                  & \multicolumn{1}{c}{Field}    & Mean                 & Pressure \\
                            &                      &                      &                          & time                 &                      & sunset               &                      &                      & \multicolumn{1}{c}{rotation} & altitude             &          \\
        \nr                 & {\AA}                & \mbox{MJD}           &                          & \mbox{UT}            & \mbox{s}             & \mbox{h}             & \mbox{J2000}         & \mbox{J2000}         & \multicolumn{1}{c}{$^\circ$} & $^\circ$             &\mbox{hPa}\\
                            &                      &                      &                          &                      &                      &                      &                      &                      &                              &                      &          \\
        \hline
                            &                      &                      &                          &                      &                      &                      &                      &                      &                              &                      &          \\
        1                   & 7700                 & 59579.05704          & 31 Dec 2021              & 01:22                & 900                  & 6.80                 & 17:52:10.6           & +70:57:17.0          & -12.31                       & 10.38                & 772      \\
        2                   & 8700                 & 59579.01002          & 31 Dec 2021              & 00:14                & 900                  & 5.67                 & 17:52:10.6           & +70:57:17.0          & 3.58                         & 9.71                 & 773      \\
        3                   & 6720                 & 59580.04950          &  1 Jan 2022              & 01:11                & 900                  & 6.62                 & 17:52:10.6           & +70:57:17.0          & -2.21                        & 10.24                & 774      \\
        4                   & 10\,500              & 59580.02506          &  1 Jan 2022              & 00:36                & 1800                 & 6.03                 & 17:52:10.6           & +70:57:17.0          & -2.22                        & 9.86                 & 774      \\
        5                   & 6720                 & 59581.05177          &  2 Jan 2022              & 01:14                & 1200                 & 6.67                 & 17:51:51.0           & +70:55:59.2          & -7.46                        & 10.42                & 775      \\
        6                   & 8700                 & 59581.03999          &  2 Jan 2022              & 00:57                & 900                  & 6.38                 & 17:52:10.6           & +70:57:17.0          & -7.46                        & 10.08                & 775      \\
        7                   & 8700                 & 59582.08453          &  3 Jan 2022              & 02:02                & 900                  & 7.45                 & 17:52:10.6           & +70:57:17.0          & -23.70                       & 11.82                & 775      \\
        8                   & 10\,500              & 60023.96559          & 20 Mar 2023              & 23:10                & 1800                 & 3.68                 & 18:12:34.2           & +59:28:00.2          & 90.00                        & 10.47                & 773      \\
        9                   & 10\,500              & 60023.93552          & 20 Mar 2023              & 22:27                & 1800                 & 2.97                 & 20:20:50.8           & +73:31:06.3          & 90.00                        & 12.90                & 773      \\
        10                  & 7700                 & 60024.98045          & 21 Mar 2023              & 23:32                & 1800                 & 4.03                 & 18:12:29.9           & +59:26:41.8          & 45.00                        & 13.01                & 772      \\
        11                  & 8700                 & 60024.93303          & 21 Mar 2023              & 22:24                & 1200                 & 2.90                 & 20:20:04.6           & +73:31:27.3          & 90.01                        & 12.68                & 772      \\
        12                  & 6720                 & 60554.95556          &  1 Sep 2024              & 22:56                & 1000                 & 3.28                 & 10:42:41.9           & +74:01:59.3          & -51.10                       & 15.04                & 772      \\
        13                  & 6720                 & 60555.03125          &  2 Sep 2024              & 00:45                & 500                  & 5.10                 & 10:42:41.9           & +74:01:59.3          & -77.70                       & 12.72                & 772      \\
        14                  & 6720                 & 60555.11736          &  2 Sep 2024              & 02:49                & 500                  & 7.17                 & 10:42:41.9           & +74:01:59.3          & -107.50                      & 14.11                & 771      \\
        15                  & 8700                 & 60554.93403          &  1 Sep 2024              & 22:25                & 1000                 & 2.77                 & 10:42:41.9           & +74:01:59.3          & 38.90                        & 15.90                & 772      \\
        16                  & 8700                 & 60555.01528          &  2 Sep 2024              & 00:22                & 1000                 & 4.72                 & 10:42:41.9           & +74:01:59.3          & 1.60                         & 12.87                & 772      \\
        17                  & 8700                 & 60555.07361          &  2 Sep 2024              & 01:46                & 1000                 & 6.12                 & 10:42:41.9           & +74:01:59.3          & 1.00                         & 12.94                & 772      \\
        18                  & 8700                 & 60555.08819          &  2 Sep 2024              & 02:07                & 1200                 & 6.46                 & 10:30:10.4           & +79:41:05.6          & 0.90                         & 18.90                & 771      \\
        19                  & 8700                 & 60555.10347          &  2 Sep 2024              & 02:29                & 1000                 & 6.83                 & 10:42:41.9           & +74:01:59.3          & 0.40                         & 13.72                & 771      \\
                            &                      &                      &                          &                      &                      &                      &                      &                      &                              &                      &          \\
        \hline
    \end{tabular}%
    }
    \tablefoot
    {\tiny Each pointing is given an ID number to allow easier comparison between Tables \ref{table:obs}, \ref{table:apparentContinuum}, and figures in Appendix \ref{app:apparent}.}
\end{table*}

\section{\label{sec:reduc}Data reduction}
The detected signal level of the continuum flux was very low, in the range 1--3\,\epx{} in all three bands where we detect a signal. For this reason, a very careful bias subtraction precise to a few tenths \epx{} was required. Before beginning the observing campaign, we observed that in a series of bias frames, the bias level shows random frame-to-frame mean level fluctuation of 1--2\,\elec{}. In addition to drifts, the bias level showed complex structure with both low- and high-frequency random modulation, and a saddle shaped settling pattern. Consequently, it was not possible to rely on stacked master bias frames since the systematic uncertainty would have surpassed the signal we tried to measure. Instead, the bias level was modelled based on the horizontal and vertical overscan regions. To correct for the the low-frequency random modulation, a non-parametric local linear regression model was fitted both row and column wise. The cross-product of the row and column models was scaled to match the mean bias level. The high-frequency random modulation was still left intact, and an additional static sinusoidal pattern were revealed. Both of these components were removed together with dark current by sampling non-illuminated areas on the detector. The adopted bias subtraction leaves a detector background that is very flat. The method was tested on a series of 120 biases which were processed using the adopted procedure, producing a mean level of $0.0\pm0.6$\,\elec{} over the entire image area. This leaves the bias subtraction itself as the main source of uncertainty in the measurement before zenith scaling, since sampling the entire slit length brings the effective readout noise to 0.3\,\elec{} in the 6720\,{\AA} band, and 0.4\,\elec{} in the rest. 

Cosmic rays were then removed with a Python implementation of the L.A.Cosmic algorithm\footnote[1]{\url{https://lacosmic.readthedocs.io/en/stable/index.html}}\citep{Dokkum01}. The dark current was sampled from the detector area not exposed to light. No flat fielding was done since the entire slit length was to be collapsed into 1D and pixel-to-pixel variations would average out. The 2D wavelength solution and rectification was made using packages in IRAF \citep{IRAF1,IRAF2}, either by using Thorium-Argon arc lamp lines (6720, 7700, 8700\,{\AA}) or sky lines if sufficient number of arc lines were not visible in the band (10\,500\,{\AA}). The sky lines were identified based on \cite{Osterbrock94} and \cite{Rousselot00} night-sky emission line atlases containing the brightest atmospheric emission lines. In the case of 6720\,{\AA} band, no spectral lines were listed in the observatory lamp line maps, and the arc lamp was first observed without NB filter to identify spectral lines with the band pass. The newly identified lines were used for fitting the 2D wavelength solution. The spectra were 2D rectified and then median collapsed to 1D. Sensitivity functions for each band were derived from the 10\arcsec -slit, or slitless standard star observations, and the spectra were calibrated. Sensitivity functions for each run were derived.

\section{\label{sec:analysis}Analysis and results}

\subsection{\label{sec:apparent}Apparent continuum radiance}
The apparent airglow continuum radiance was measured from regions of the spectra close to the peak transmission of the NB filter where no sky lines are reported in line lists of \cite{Hanuschik03} for the 6720, 7700 and 8700\,{\AA} bands, and in \citep{Rousselot00} for the 10\,500 {\AA} band. These lines lists are not complete, but they do contain the brightest atmospheric emission lines. We indicate \citet{Loo07,Loo07erratum} computed OH lines as reference in our figures, but since we cannot identify the lines in our spectra, we do not reject these regions from our analysis. Einstein coefficients for the transitions are generally low and the lines can be expected to faint. However, we cannot rule out the possibility of being affected by them. We reject a region corresponding to 95\% of the LSF (see Sec. \ref{sec:lsf}) around each skyline found in the \cite{Hanuschik03} or \cite{Rousselot00} line lists. Weighted means were calculated from the accepted wavelength bins and are considered as the apparent continuum radiance for the band pass. Additionally, the selected regions were visually inspected not to contain spectral lines. We observe the apparent non-corrected airglow continuum radiance at large zenith distances to be on average 350, 560, 660\,\phs, or 20.92, 20.26, 19.95\,\magarcsec in the 6720, 7700 and 8700\,{\AA} respectively, and we derive an upper limit 1500\,\phs of 20.8\,\magarcsec on the 10\,500\,{\AA} band. The apparent observed flux calibrated spectra are presented in Figs.~\ref{fig:app1}, \ref{fig:app2}, \ref{fig:app3}, \ref{fig:app4}, \ref{fig:app5}, \ref{fig:app6}, and \ref{fig:app7} for all four band passes. We present the apparent continuum radiance in Table \ref{table:apparentContinuum} alongside the zenith scaled values.

\subsection{\label{sec:zenithscaling}Zenith equivalent radiance}
In order to allow comparison with works in the literature, the zenith normalized continuum radiance is calculated and reported in Table \ref{table:apparentContinuum}. Due to ZL and airglow being emitted from a large angular area, their effective airmass and optical depth scale differently than that of a point source (see Sec. \ref{sec:scaling}). For the most part we follow \cite{Noll12} in the zenith scaling, adapting their model to our case. For the scaling purposes, we assume the airglow continuum to originate from the same atmospheric layer as the \mbox{OH} line emission. We take {OH} layer altitude and thickness from rocket borne experiments \citep{Lopez-Moreno87,Baker88}, and assume that the airglow continuum emission originates from a mean altitude of 87\,km with a layer full-width-at-half-maximum of 9\,km. The following steps were taken in the zenith normalization process:

\begin{enumerate}
    \item Sky spectra are reduced, flux calibrated, and the apparent interline continuum is measured at the observed zenith distance $z$.
    \item Contribution of extra-atmospheric emission is computed for the time and pointing.
    \item Effective optical depths for airglow and ZL are calculated.
    \item Extra-atmospheric emission is attenuated and subtracted from the total observed intensity.
    \item A scaling factor for the thickness of the airglow emitting layer is calculated, and line-of-sight emitted airglow is scaled to zenith equivalent.
    \item Airglow emission scattered to the line-of-sight at zenith is added to give apparent zenithal airglow radiance.
\end{enumerate}

\noindent
The assumption of atmospheric attenuation being due to scattering only is generally true for the 6720, 8700 and 10\,500\,{\AA} bands, but the blue half of the 7700\,{\AA} band is affected by \mbox{O$_2$} absorption.

\subsection{\label{sec:scaling}Airmass scaling}
The altitude and thickness of the emitting airglow continuum layer is assumed to be similar to that of OH line emission \citep{Baker88}. Due to the large zenith distance $z$ observed, and the fact that the emitting airglow layer has a finite thickness, the airglow emission $I_{\rm ag}$, the typical plane parallel atmosphere airmass $X = \sec(z)$ scaling cannot be used. Thickness of the emitting airglow layer increases significantly less steeply, and the typical relation to describe the scaling $s_{\rm ag}$ is the so called van Rhjin function,

\begin{align}
    \label{eq:vanrhijn}
    s_{\rm ag} = \frac{I_{\rm ag}(z)}{I_{\rm ag}(0)} =& ~\left(1 - \left( \frac{R\sin(z)}{R+h} \right)^2 \right)^{-1/2}
\end{align}
\newline
\noindent
where $I_{\rm ag}(z)$ is airglow intensity at zenith distance $z$, $I_{\rm ag}(0)$ is airglow intensity at zenith, $R$ is the radius of Earth, and $h$ the height of the emitting layer above Earth's surface. Eq. \ref{eq:vanrhijn} makes the assumption that the thickness of the emitting layer can be neglected which is generally valid for lower $z$, but breaks down when either the observer gets closer to the emitting layer, or points close to the horizon. We use an earlier step of van Rhijn's derivation and the airglow layer thickness is taken as a side length difference of two scalene obtuse triangles, which is the same as Eq. 19 in \cite{vanRhijn21}. Using positive quadratic solutions for both triangles, the OH emission scaling $s_{\rm ag}$ factor becomes 

\begin{align}
    \label{eq:obstusetriangle}
    s_{\rm ag} = ~\frac{I_{\rm ag}(z)}{I_{ag}(0)} =& \sqrt{R^2cos^2(z)+2Rh_2+h_2^2} \\ 
    &- \sqrt{R^2 cos^2(z)+2Rh_1 + h_1^2}, \nonumber
\end{align}
\newline
\noindent
where $z$ is the zenith distance, $R$ is Earth's radius, $h_1$ is the altitude of the emitting layer, and $h_2$ the altitude of the observer. Our observations cover zenith distances of 77 -- 81\degr, and the observer altitude begins to make small difference when $z\gtrsim79$\degr.

Due to scattering to the line-of-sight the airmass of a large solid angle source scales differently than that of a point source. For diffuse emission outside of the atmosphere, we take

\begin{align}
    \label{eq:Xzl}
    X_{\rm zl}(z) =& ~( 1 - 0.96 \sin^2(z)  )^{-1/2}
\end{align}
\newline
\noindent
\citep{Krisciunas91,Noll12} which we use as the ZL airmass. Light from the airglow layer travels a shorter distance in the atmosphere, and we use  

\begin{align}
    \label{eq:Xag}
    X_{\rm ag}(z) =& ~( 1 - 0.972 \sin^2(z) )^{-1/2}
\end{align}
\newline
\noindent
as the airglow airmass \citep{Noll12}. Also, in order to scale the airglow emission to zenith, optical depth reduction due to the line-of-sight scattering needs to be considered. We follow the treatise of \cite{Noll12}, and apply their methodology to our site and extend their parametrization to our case.

Optical depths for large solid angle sources are reduced due to the line-of-sight scattering. The reduction can be presented as an effective optical depth \citep{Noll12}, 

\begin{align}
    \tau_{\rm eff} = ~f_{\rm ext}\tau_0,
\end{align}
\newline
\noindent
where $f_{\rm ext}$ is the extinction reduction factor, and $\tau_0$ zenithal optical depth. We take the \cite{Noll12} reduction factors $f_{\rm ext}$ for ZL Rayleigh and Mie scattering,

\begin{align}
    f_{\rm ext,\,zl,\,R} =& ~1.407\log I_{\rm zl} - 2.692 \\
    f_{\rm ext,\,zl,\,M} =& ~1.309\log I_{\rm zl}  - 2.598
\end{align}
\newline
\noindent
where $I_{zl}$ is the ZL outside atmosphere in units of \mbox{$\mathrm{W^{-8}\,m^{-2}\,\mu m^{-1}\,sr^{-1}}$} with requirement that \mbox{$\log I_{\rm zl}\leq2.44$}, a criterion met in all our observations. Similarly, we take reduction factors for the airglow emission Rayleigh and Mie components as,

\begin{align}
    f_{\rm ext,\,ag,\,R} =& ~1.669\log X_{\rm ag} - 0.146 \\
    f_{\rm ext,\,ag,\,M} =& ~1.732\log X_{\rm ag} - 0.318
\end{align}
\newline

\noindent
Components contributing to the optical depth are Rayleigh scattering, Mie scattering, and molecular absorption and the zenithal optical depth is

\begin{align}
    \tau_0(\lambda) =& ~\tau_R(\lambda) + \tau_M(\lambda) + \tau_A(\lambda),  
\end{align}
\newline
\noindent
where the $\tau_R$, $\tau_M$, and $\tau_A$ stand for the Rayleigh, Mie, and absorption components respectively. $\tau_M$, and $\tau_A$ are estimated based on ESO Sky Calc model for La Silla. \cite{Liou02} gives the following formula for Rayleigh scattering optical depth with wavelengths in \mum{},

\begin{align}
    \label{eq:rayleigh}
    \tau_{R}(\lambda, h, p) =& ~\frac{p}{p_s}\left(a+bh\right)\lambda^{-(c+d\lambda + e/\lambda)},
\end{align}
\newline
\noindent
where, $h$ is altitude in kilometers, $p$ is pressure, \mbox{$p_S=1013.25$\,hPa}, and the constants are \mbox{$a=0.00864$}, \mbox{$b=6.5\times10^{-6}$}, \mbox{$c=3.916$}, \mbox{$d=0.074$}, and \mbox{$e=5\times10^{-2}$}. Eq. \ref{eq:rayleigh} provides comparable though slightly lower $\tau_R$ values than the non-pressure dependent \cite{King85} model for Observatorio del Roque de los Muchachos. 

Atmospheric transmission as a function of wavelength is,

\begin{align}
    t(\lambda, z) =& ~e^{-\tau_0 (\lambda)X_{}(z)}
\end{align}
\newline
\noindent
where $\tau_0$ is optical depth in zenith, and $X$ is the airmass. We calculate effective transmissions $t_{\rm ag}$ and $t_{\rm zl}$ for airglow and ZL respectively.

Finally, considering the different airmasses, and the optical depth reduction factors for ZL and airglow, the zenith equivalent airglow intensity can be calculated as,

\begin{align}
    \label{eq:Iag0}
    I_{\rm ag}(0) =& ~\frac{I_{\rm ag}(z) - I_\star - (I_{\rm zl} + I_{GBL} + I_{EBL})\,t_{\rm eff,\,zl}(\lambda, z)}{t_{\rm eff,\,ag}(\lambda, z)\,s_{\rm ag}} \, t_{\rm eff, ag}(\lambda, 0)
\end{align}
\newline
\noindent
where $I_{\rm ag}(z)$ is the observed airglow intensity at zenith distance $z$, $I_\star$ scattered star light, $I_{\rm GBL}$ galactic background light, and $I_{\rm EBL}$ extra-galactic background light intensities, $t_{\rm eff, ag}$ and $t_{\rm eff, zl}$ are the effective transmission for airglow and ZL, and $s_{\rm ag}$ the scaling factor to compensate for the difference in apparent emitting layer thickness. Notice that $t_{\rm eff,ag}(\lambda,0)$ is greater than one due to the line-of-sight scattering. We calculated the mean airmasses for each observation from the telescope altitude pointing trajectory during the observation.

\subsection{Uncertainty due to extinction}
Observing at zenith distance $z\sim80\degr$, the line of sight extends several hundreds of kilometers in the lower atmosphere before reaching altitude of mesopause. The short wavelength side of the 7700\,{\AA} band pass contained significant \mbox{O$_2$} absorption. Instead of trying to correct for the absorption, we ignored the affected range. None of our bands are affected by water vapor absorption. The largest airmass data point for \cite{Noll12} $f_{\rm ext\, ag}$ fit is $X=2.75$. We extrapolate their fit to $X>4$. Overestimating $f_{\rm ext,\,ag}$ leads to overestimating $I_{\rm ag}(0)$.

\subsection{\label{sec:zl}Zodiacal light}
The ZL spectrum is a reflected solar spectrum from inter-planetary dust, and the ZL surface brightness depends on the solar elongation angle, and longitude of Earth's ascending node. The state-of-art ZL models are based on COBE/DIRBE SWIR data \citep{Kelsall98,Wright98}. The bluest observed COBE/DIRBE band is centered at 1.25\,\mum{}, and the DIRBE ZL models are extrapolated to VIS--NIR wavelengths. Extrapolation assumes an albedo for inter-planetary dust which is not well known, leaving non-negligible uncertainty on the modeled ZL radiance at VIS and NIR range. The \cite{Wright98} model enforces, so called 'strong no-zodi principle' requiring the ZL at high galactic latitude at 25\,\mum{} to be isotropic and constant in time. Due to the condition, \cite{Wright98} gives systematically higher ZL radiance than the \cite{Kelsall98} model. It has been suggested that the \cite{Kelsall98} ZL model underestimates the ZL for wavelengths shorter than $<3.5$\,\mum{} \cite{Tsumura13}, and that the model may miss a diffuse ZL component \citep{Kawara17}. It does not seem to be settled in the literature which model to adopt for VIS and NIR wavelengths, and since the zenith scaling in this work is ZL model dependent, we computed results based on both models. Additionally, we compared against \mbox{ESO SkyCalc v.2.0.9} \footnote[1]{\url{https://www.eso.org/observing/etc/bin/gen/form?INS.MODE=swspectr+INS.NAME=SKYCALC}} \citep{Noll12,Jones13}ZL model which is based on visual wavelength observations of \cite{Levasseur80} on Tenerife in the late 1960's and early 1970's, and uses the reddening relations of \cite{Leinert98}.

We find that there is few a percentage difference between the scaled zenithal airglow radiance if we use DIRBE \cite{Kelsall98} ZL model or ESO SkyCalc \cite{Noll12} ZL model, which is below the uncertainty of our measurement. Using \cite{Wright98} DIRBE ZL model with \citep{Gorjian00} parameters, our results ends up $\sim$10\% lower compared to the \cite{Kelsall98} DIRBE model. We report and show $I_{\rm zl}$ values based on the \cite{Kelsall98} model only. We have calculated the DIRBE ZL radiance for the date and pointing with InfraRed Science Archive (IRSA), Infrared Processing \& Analysis Center (IPAC) Euclid background model calculator Versions 1 and 4 \footnote[2]{\url{https://irsa.ipac.caltech.edu/applications/BackgroundModel/}}, the Version 1 being based on \cite{Kelsall98} model, and the Version 4 on the \cite{Wright98, Gorjian00} model.

\subsection{Other diffuse radiation}
As noted in the Section \ref{sec:fields}, we assume that the Moon does not contribute to the total observed sky radiance. In addition to the ZL, we consider contribution from scattered star light, and galactic- ($I_{\rm GBL}$), and extra-galactic background light ($I_{\rm EBL}$) which are very low compared to the ZL. For the scattered star light surface brightness $I_\star$ we use values of 7, 4.4, 3.1, and 2.3\,\phs{} for 6720, 7700, 8700 and 10\,500\,{\AA} bands respectively \citep{Noll12}. We take $I_{\rm GBL}$, and $I_{\rm EBL}$ components from the same IRSA/IPAC calculator which is based on \cite{Arendt98}. Apart from few pointings, both $I_{\rm GBL}$ and $I_{\rm EBL}$ contribute negligible $\sim$1\,\phs{} in the band passes.

\begin{table*}
    \caption{\label{table:apparentContinuum}Apparent observed and inferred zenith equivalent airglow continuum radiance.}
    \centering
    \begin{tabular}{ccc@{\hskip 0.6pt}c c@{\hskip 0.6pt}c c@{\hskip 0.6pt}c c@{\hskip 0.6pt}c}
    \hline\hline
            &       &       &                         &      &       &              &                         &               &                           \\
    Band    &  ID   & \multicolumn{2}{c}{Total continuum}         &\multicolumn{2}{c}{Zodiacal light} &\multicolumn{2}{c}{Airglow continuum}   &\multicolumn{2}{c}{Airglow continuum}\\
            &       &\multicolumn{2}{c}{\textit{apparent observed}} &\multicolumn{2}{c}{\textit{\cite{Kelsall98}}}&\multicolumn{2}{c}{\textit{zenith, emitted}} & \multicolumn{2}{c}{\textit{zenith, emitted + scattered}} \\
            &       &              &                         &      &       &              &                         &               &                          \\
    {\AA}   & \nr   &${\rm ph\,s}^{-1}{\rm m}^{-2}{\rm \mu m}^{-1}$ & mag & ${\rm ph\,s}^{-1}{\rm m}^{-2}{\rm \mu m}^{-1}$ & mag &  ${\rm ph\,s}^{-1}{\rm m}^{-2}{\rm \mu m}^{-1}$ & mag  &   ${\rm ph\,s}^{-1}{\rm m}^{-2}{\rm \mu m}^{-1}$ & mag \\
            &       &${\rm arcsec}^{-2}$ & ${\rm arcsec}^{-2}$ & ${\rm arcsec}^{-2}$ & ${\rm arcsec}^{-2}$ & ${\rm arcsec}^{-2}$ & ${\rm arcsec}^{-2}$ &  ${\rm arcsec}^{-2}$ & ${\rm arcsec}^{-2}$ \\
            &       &              &                         &      &       &              &                         &              &                          \\
    \hline
            &       &              &                         &      &       &              &                         &              &                          \\
    6720    & 3     & 271 $\pm99$  & 21.20$^{+0.49}_{-0.34}$ &  59  & 22.85 &   53 $\pm19$ & 22.97$^{+0.48}_{-0.33}$ &  73  $\pm27$ & 22.62$^{+0.50}_{-0.34}$ \\
    6720    & 5     & 255 $\pm75$  & 21.26$^{+0.38}_{-0.28}$ &  59  & 22.85 &   48 $\pm14$ & 23.08$^{+0.37}_{-0.28}$ &  67  $\pm20$ & 22.71$^{+0.38}_{-0.28}$ \\
    6720    & 12    & 525 $\pm32$  & 20.48$^{+0.07}_{-0.06}$ &  64  & 22.76 &   78 $\pm5$  & 22.55$^{+0.07}_{-0.07}$ &  107 $\pm7$  & 22.20$^{+0.07}_{-0.07}$ \\
    6720    & 13    & 440 $\pm64$  & 20.67$^{+0.17}_{-0.15}$ &  64  & 22.76 &   74 $\pm11$ & 22.61$^{+0.17}_{-0.15}$ &  102 $\pm15$ & 22.26$^{+0.17}_{-0.15}$ \\
    6720    & 14    & 316 $\pm64$  & 21.03$^{+0.25}_{-0.20}$ &  64  & 22.76 &   46 $\pm9$  & 23.12$^{+0.24}_{-0.19}$ &  63  $\pm13$ & 22.78$^{+0.25}_{-0.20}$ \\
            &       &              &                         &      &       &              &                         &              &                         \\
    7700    & 1     & \la 560      & \la 20.26               &   49 & 22.91 &   \la 92     & \la 22.22               &  \la 126     & \la 21.88               \\
    7700    & 10    & \la 560      & \la 20.26               &   50 & 22.88 &   \la 79     & \la 22.39               &  \la 109     & \la 22.04               \\
            &       &              &                         &      &       &              &                         &              &                         \\
    8700    & 2     & 536 $\pm216$ & 20.18$^{+0.56}_{-0.37}$ &   41 & 22.97 &   80 $\pm32$ & 22.24$^{+0.55}_{-0.37}$ &  110 $\pm44$ & 21.89$^{+0.55}_{-0.37}$ \\
    8700    & 6     & 914 $\pm233$ & 19.60$^{+0.32}_{-0.25}$ &   41 & 22.97 &  139 $\pm35$ & 21.64$^{+0.31}_{-0.24}$ &  191 $\pm49$ & 21.30$^{+0.32}_{-0.25}$ \\
    8700    & 7     & 434 $\pm242$ & 20.40$^{+0.89}_{-0.48}$ &   41 & 22.97 &   58 $\pm43$ & 22.59$^{+1.47}_{-0.60}$ &   79 $\pm59$ & 22.25$^{+1.49}_{-0.61}$ \\
    8700    & 11    & 510 $\pm199$ & 20.23$^{+0.54}_{-0.36}$ &   43 & 22.91 &   64 $\pm25$ & 22.48$^{+0.54}_{-0.36}$ &   88 $\pm34$ & 22.14$^{+0.53}_{-0.35}$ \\
    8700    & 15    & 710 $\pm196$ & 19.87$^{+0.35}_{-0.26}$ &   45 & 22.87 &   88 $\pm24$ & 22.14$^{+0.35}_{-0.26}$ &  120 $\pm33$ & 21.80$^{+0.35}_{-0.26}$ \\
    8700    & 16    & 887 $\pm222$ & 19.63$^{+0.31}_{-0.24}$ &   45 & 22.87 &  120 $\pm30$ & 21.80$^{+0.31}_{-0.24}$ &  165 $\pm41$ & 21.45$^{+0.31}_{-0.24}$ \\
    8700    & 17    & 751 $\pm248$ & 19.81$^{+0.44}_{-0.31}$ &   45 & 22.87 &  100 $\pm33$ & 22.00$^{+0.43}_{-0.31}$ &  138 $\pm46$ & 21.65$^{+0.44}_{-0.31}$ \\
    8700    & 18    & 563 $\pm194$ & 20.12$^{+0.46}_{-0.32}$ &   42 & 22.94 &   66 $\pm23$ & 22.45$^{+0.47}_{-0.32}$ &   90 $\pm31$ & 22.11$^{+0.46}_{-0.32}$ \\
    8700    & 19    & 618 $\pm255$ & 20.02$^{+0.58}_{-0.38}$ &   45 & 22.87 &   79 $\pm33$ & 22.25$^{+0.59}_{-0.38}$ &  109 $\pm45$ & 21.90$^{+0.58}_{-0.38}$ \\
            &       &              &                         &      &       &              &                         &              &                         \\
    10\,500$^{\dagger}$ & 4 & \la 3153  & \la 18.05           &   31 & 23.07 &      \la 402 & \la 20.28               & \la 553      & \la 19.94               \\
    10\,500$^{\dagger}$ & 8 & \la 1922  & \la 18.58           &   32 & 23.03 &      \la 240 & \la 20.84               & \la 330      & \la 20.50               \\
    10\,500$^{\dagger}$ & 9 & \la 1508  & \la 18.85           &   32 & 23.03 &      \la 181 & \la 21.15               & \la 249      & \la 20.80               \\
            &       &             &                          &      &       &              &                         &              &                         \\
    \hline
    \end{tabular}
    \tablefoot
    {\tiny Apparent airglow continuum radiance at high airmass reported without any corrections. \emph{zenith, emitted + scattered} corresponds directly to Eq.\ref{eq:Iag0}. ID can be compared against table of observations, Table \ref{table:obs}. \tablefoottext{$\dagger$}{No detection, upper limit.}
    }
\end{table*}

\begin{figure*}[ht]
    \begin{center}
    \resizebox{\hsize}{!}{\includegraphics[width=\textwidth]{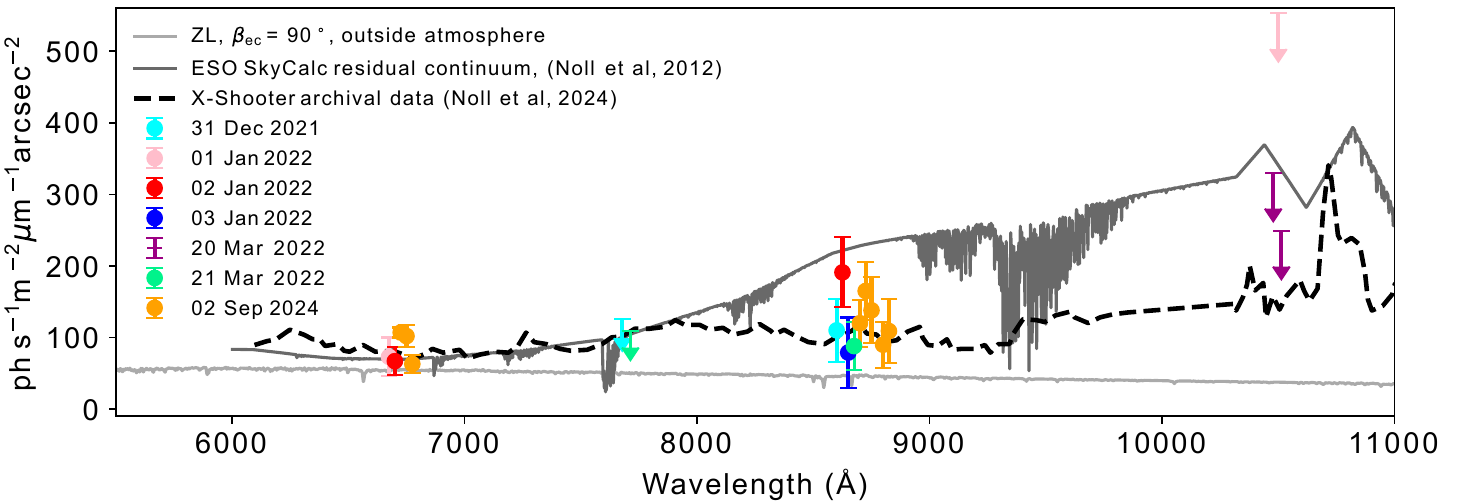}}
    \end{center}
    \caption{\label{fig:continuumAll}Zenith equivalent airglow continuum radiance in the observed band passes according to Eq. \ref{eq:Iag0}. Data points are shifted by 25\,{\AA} steps to prevent them from being plotted over each other. ESO SkyCalc yearly average residual airglow continuum \citep{Noll12,Jones13}, X-shooter 2009--2019 mean airglow continuum \citep{Noll24}, and ZL spectrum above atmosphere towards ecliptic north pole (\cite{Meftah18} solar spectrum scaled to \cite{Kelsall98} model). Upper limits are derived in 10\,500\,{\AA} band are indicated by arrows.
    }
\end{figure*}

\begin{figure}[ht]
    \begin{center}
    \resizebox{\hsize}{!}{\includegraphics[width=\hsize]{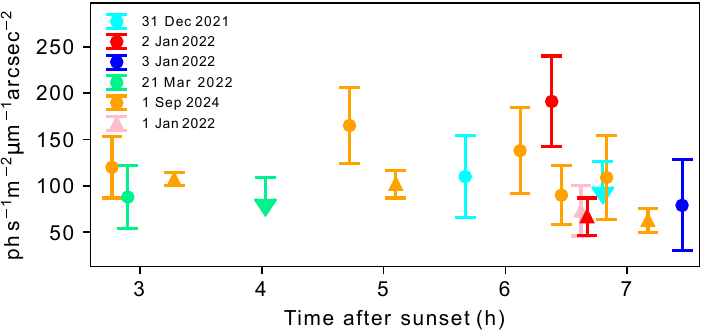}}
    \end{center}
    \caption{\label{fig:aftersunset}Zenith equivalent continuum radiance as a function of time after sunset in 6720\,{\AA} (triangles), 7700\,{\AA} (upper limits, arrows down), and 8700\,{\AA} (circles) bands.
    }
\end{figure}

\section{\label{sec:discussion}Discussion}

\subsection{Airglow or instrumental}
This work originally set out to investigate if the previously observed airglow continuum values can be explained by grating scattered light contribution. At the observed wavelengths, the detected continuum signal is clear despite being faint, and its presence cannot be attributed to the grating scattered light originating from the brighter OH lines within our bandwidths (see Sec.\ref{sec:lsf}). We estimate conservatively the spectrograph LSF wing contribution to be <\,1\,\phs{} in the 6720\,{\AA} band, and <\,10\,\phs{} in 7700 and 8700\,{\AA} bands. The total grating scattered light contribution is below the systematic uncertainty limits: the observed continuum is not introduced by the grating scattered light in our optical system. The observed continuum or pseudo-continuum is real, and  of atmospheric or extra-atmospheric origin. With our R$\sim$4000 spectra we will not be able distinguish between true continuum and densely spaced weaker atmospheric spectral lines. Although unidentifiable, our chosen line free regions within our band passes may contain OH lines. We indicate positions computed by \cite{Loo07,Loo07erratum} in our figures.

\subsection{Unknown and unidentified lines}
The spectral ranges that are considered to be devoid of atmospheric spectral lines are based on the observational line list of \cite{Hanuschik03} and \cite{Rousselot00}. The selected regions contain additional known OH transitions \citep{Loo07,Loo07erratum} which we do not detect in our spectrum. In the 8700\,Å band we see correlation between the continuum and OH and O$_2$ line emission, for which an explanation is that we are affected by the fainter OH lines which we are unable to identify. Several of these OH transition have long life times, and the lines can be expected to be faint. We indicate the line with tick marks scaled by the line's Einstein coefficient in our figures. Additionally, a faint line is detected in all spectra observed during the New Year 2022 run in range 8731 - 8732\,{\AA}, which disappears from rest of the spectra. The following ions can be found in NIST line data base \ion{Ti}{iii}, \ion{W}{i}, \ion{Fe}{vii} \ion{Mn}{ii}, and \ion{V}{ii} \citep[NIST,][]{NIST_ASD}, all of them odd appearances on the night sky spectrum. The line could be potentially originating from either fireworks, or deposition of micrometeorites \citep{Plane15}. These observations were carried out two weeks after the peak of Geminides meteor shower. Due to the lack of certainty on its origin and its faintness leads us not to reject the spectral range in our analysis.

\subsection{Temporal variability and color}
In a typical case, we observed several hours after the sunset. However, based on the few measurements at the beginning of night, the airglow continuum radiance does not seem to change much over the course of the night (Fig. \ref{fig:aftersunset}). By and large, the measured continuum seems stable, and is almost flat in color with potential slight increase towards red. Typically, when we measured two bands during the same night, the bands measured a comparable level, the only exception being 1 Jan 2022 which shows significant difference in emission between 6720 and 8700\,{\AA} bands (see Sec. \ref{sec:solarrel} for further discussion). The 7700 and 8700\,{\AA} bands were observed with a one hour time difference on 30 Dec 2021 and 21 Mar 2022. On both occasions, the 7700\,{\AA} band measures equal radiance to 8700\,{\AA} band. The 6720 and 8700\,{\AA} bands were observed observed similarly with somewhat shorter time difference on 2 Jan 2022, and several times on 1 Sep 2024. On 1 Sep 2024, the 8700\,{\AA} radiance stays constant within our measurement errors with a weighted average 126\,\phs{}, or 21.74\,\magarcsec{} with a slight tendency for the radiance to decrease towards the end of night. The 6700\,{\AA} band has lower measurement errors and shows statistically significant decrease over the night with a $\sim$40\,\% drop in radiance between the first and last measurement. Color of the resulting continuum spectrum is flat with a marginal increase towards the red, see Fig. \ref{fig:continuumAll}. Our data points overlap with the mean airglow spectrum of \cite{Noll24} for Cerro Paranal, Chile.

\subsection{\label{sec:solarrel}Relation to solar activity}
\begin{figure*}[!ht]
    \begin{center}
    \includegraphics[width=\textwidth]{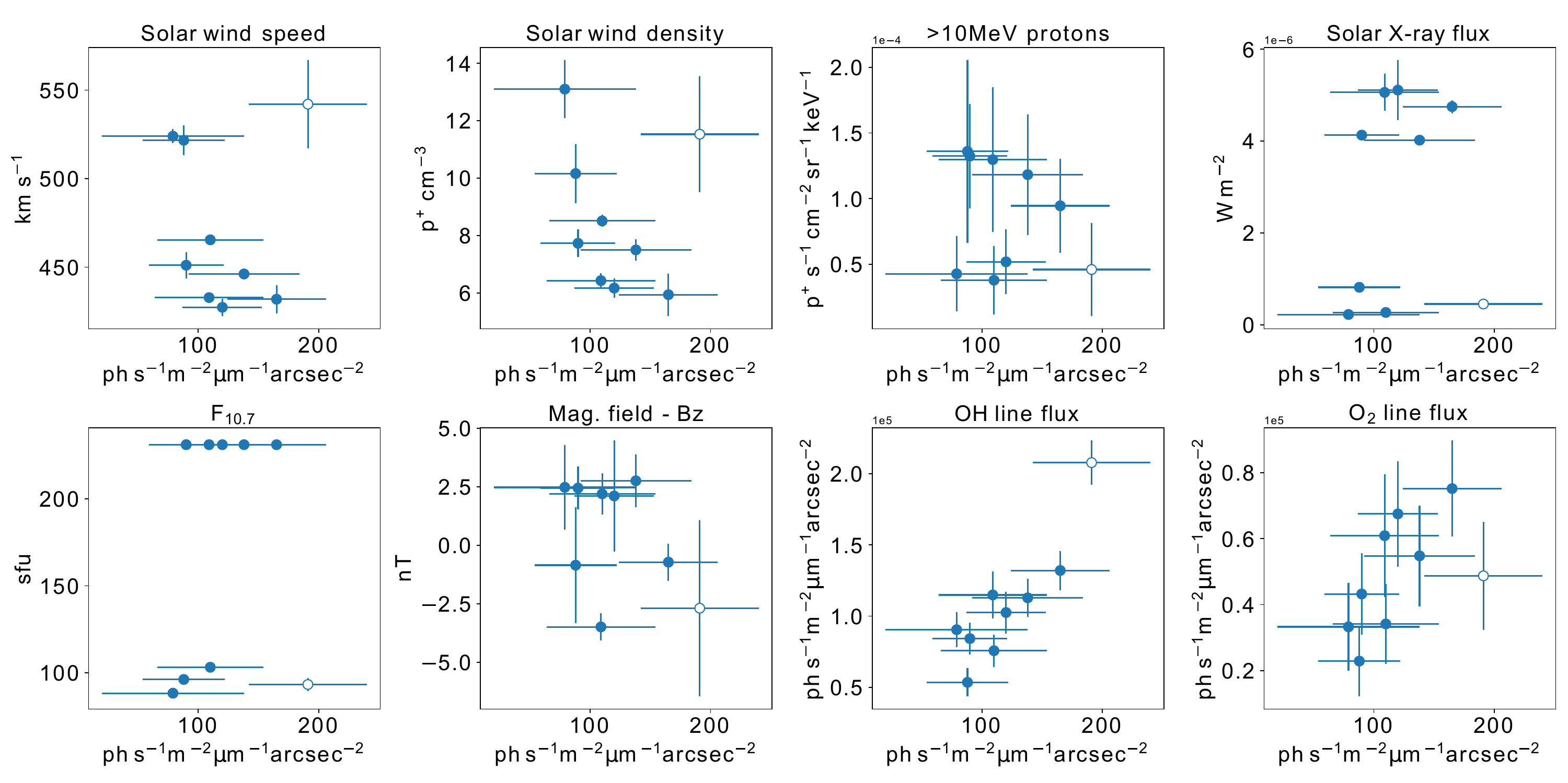}
    \end{center}
    \caption{\label{fig:solarcorr}The airglow continuum dependence in the 8700\,{\AA} band pass on Solar activity parameters, and OH and O$_2$ line emission. High solar activity has co-coincided with slow solar wind in our data set. Observation \nr\,6, 2 Jan 2022, with the highest measured 8700\,{\AA} radiance, is a potential outlier as discussed in Sec.\ref{sec:solarrel} and is marked with a hollow symbol. Presented OH flux is a sum of blended 7-3\,R \mbox{OH} lines ($\sim$8761, 8768, 8778, and 8791\,{\AA} in our spectra), while \mbox{$\mathrm{O_2}$} flux is a sum of blended \mbox{$\mathrm{O_2}$} P and Q -branch lines in the range 8645--8696\,\AA.
    }
\end{figure*}

On 2 Jan 2022 the airglow continuum and OH line radiance in the 8700\,{\AA} band pass (ID 6) measured elevated values respective to the other nights prompting us to look for an explanation. We seek relation to various Solar activity parameters (Fig.\ref{fig:solarcorr}). Unfortunately, the data set is small and biased in respect to the space weather: strong solar radiation activity has coincided with slow solar wind conditions. No clear conclusion can be drawn given the high temporal variability of the airglow radiance. The 2 Jan 2022 the 8700\,{\AA} (ID 6) measures the highest level in our data, and standing out as a possible outlier in the solar activity plots (hollow marker in Fig. \ref{fig:solarcorr}). Potentially, the increase in radiance could be due to a coronal mass ejection hitting Earth's magnetosphere during the observation. A C1.06 -class flare has taken place 45~min before the beginning of observation, and another B8.42 flare 30~min before the observation. At the beginning of the exposure, DSCOVR registered an increase in solar wind speed from average 525\,\mbox{$\mathrm{km\,s^{-1}}$} up to 600~\mbox{$\mathrm{km\,s^{-1}}$}, and solar proton density increse from average 11~$\mathrm{p^+\,cm^{-1}}$ to momentarily >16~$\mathrm{p^+\,cm^{-1}}$, which seems like the coronal mass ejection passing DSCOVR. It would have taken about 40--45\,min for the particles reach to Earth's magnetosphere, which would match the coronal mass ejection hitting right at the beginning of the observation. The 6720\,{\AA} band (ID 5) observed right after the 8700\,{\AA} measures a low continuum level, but shows elevated \mbox{OH 6-1~P2$_{\rm e,f}(6.5)$} and \mbox{6-1~P1$_{\rm e,f}(7.5)$} line emission considering that the observation was done over 6.5~h after sunset.
 
\subsection{\label{sec:comparison}Comparison to the literature}
We attempted an exhaustive literature search as a part of the work to allow comparison of measurements over previous solar cycles. However, direct comparison deemed to be difficult due to differences in reporting.  Few publications report the total observed continuum radiance \citep{Broadfoot68,Maihara93b,Cuby00,Hanuschik03,Sullivan12,Ellis12,Trinh13,Oliva15,Nguyen16}, while others report zenith scaled equivalent or a breakdown by component \citep{Sternberg72,Gadsden73,Noxon78}, which is ZL model and atmospheric scattering dependent. Some cases exact pointing and time of observation were not reported. Allowing differences in reporting, the airglow continuum levels measured previously can be viewed comparable to what we find in this work. Previous airglow continuum measurements have been collected in Table \ref{table:previous}, and units from each other study have been converted to the units used in our work.

While OH line emission has been studied at Observatorio del Roque de los Muchachos \citep{Oliva13,Franzen18,Franzen19}, only a single report on NIR or SWIR airglow continuum can be found in literature \citep{Oliva15}. \cite{Oliva15} find comparable continuum radiance in H-band that has been reported for Las Campanas, Chile \citep{Sullivan12}, Cerro Paranal, Chile \citep{Noll24}, Mauna Kea, Hawaii \citep{Maihara93b}, and Siding Spring, Australia \citep{Trinh13}. All VIS--NIR continuum radiance measurements have been done elsewhere, recently, most notably at Cerro Paranal, Chile \cite{Hanuschik03, Noll24}. We will compare our results to measurements made on other sites.

\cite{Hanuschik03} observes apparent continuum radiance of 300--350\,\phs{} in the wavelength range observed in this work. Exact dates of observations are not reported, but the reported Moon illumination fraction restricts the dates to 20 - 21 June 2001. \cite{Hanuschik03} observes at mean airmass of $z=1.11$, forcing the pointings quite close to ecliptic equator with maximum distance being approximately $\beta{\rm ec}\sim-33\degr$. The \cite{Kelsall98} model gives $\sim$100\,\phs{} ZL contribution to the total apparent continuum. The \cite{Hanuschik03} spectra appears bluer than we find in this work, but the level is comparable after taking ZL into account to the nights associated with stronger solar activity in our study. On 20 - 21 June 2001 DRAO $F_{10.7}$ ranged between 199--204\,sfu indicating moderately high solar activity.

\cite{Sternberg72} report a time series of ZL corrected observations at 8200\,{\AA} towards the North Celestial Pole from Haute-Provence observatory, during nights 18 -- 19 Aug 1969 with a nightly decrease from 300 to 190\,\phs{}. \cite{Kelsall98} ZL radiance for their pointing is $\sim$40\,\phs{}. The level is similar to found by and \cite{Hanuschik03}. The decay resembles behavior observed in this work on 1 Sep 2024. \cite{Sternberg72} notes bad weather potentially affecting the observations, but does not elaborate. \cite{Sternberg72} finds a close to constant continuum emission of 75\,\phs{} at 6750\,{\AA} on 2 June 1970.

ESO Sky Calc \citep{Noll12,Jones13, Noll13-vlt-man} residual airglow continuum offers another point of comparison. The residual airglow continuum model is based on 26 averaged early X-shooter spectra. The \cite{Noll12} residual continuum shows significant increase towards red between 7000 and 9000\,{\AA} which resembles to \cite{Broadfoot68}. Similar reddening is also reported by \cite{Gadsden73}. \cite{Noxon78} did not observe radiance to increase and suggests that \cite{Gadsden73} and \cite{Broadfoot68} reddening may have been instrumental. More recently, \cite{Noll24} attributes the \cite{Noll12} increase towards longer wavelengths to the low resolution of FOcal Reducer and low dispersion Spectrograph 1 (FORS 1) used in the \cite{Noll12} study. Nonetheless, we may have witnessed similar reddening during the night of 1 Sep 2024 associated with stronger solar activity. Recently, \cite{Noll24} finds much a flatter airglow continuum spectrum based on a larger X-shooter dataset, covering from late 2009 to 2019. The flat continuum spectrum is similar to what we typically observed at Roque de los Muchachos.

\subsection{\label{sec:implications}Implications}
Our observations indicate a dark interline NIR sky. To take the full advantage of it for faint object spectroscopy, its properties should be investigated further. Our work may also encourage introduction of OH line suppressors and grating stray light reducing spectrograph designs to workhorse NIR and SWIR spectrographs. Successful OH line suppressors based one line masks \citep[e.g.][]{Maihara93a, Iwamuro01, Parry04}, and Fiber Bragg Gratings \citep[e.g.][]{Ellis12}, have been demonstrated in the literature. The grating scattering wing reducing spectrograph designs use the grating in a double-pass configuration combined with a secondary slit between the dispersions. The double-pass spectrograph can be implemented either, as a scanning monochromator \citep{Enard82} (compatible with long slits, but records only a single spectral element at a time), or as a white-pupil spectrograph with a beam rotation between the two dispersions \citep{Andersen92, Viuho22} (confined to use with round apertures, i.e. fibers, with advantage of recording full spectral range at once). 

We observed variations of up to factor of two in the VIS-NIR airglow continuum on timescales from tens of minutes to hours. Our observations do not temporally resolve the shorter timescale variations associated with the mesospheric buoyancy waves, and it is possible that the airglow continuum exhibits similar variation as the airglow line emission with minute timescales. We do not have data to study seasonal variability reported by \cite{Patat08} and \cite{Noll24}, and it is difficult to secure observing time for a non-standard optical configuration for a long term monitoring program. Temporal variability of the continuum radiance needs to be taken into account when designing faint object spectrographs for ground-based observation even when OH suppressing or a double-pass techniques are being used.

\section{\label{sec:conclusion}Conclusion}
We have measured the NIR airglow continuum radiance at the Observatorio del Roque de los Muchachos with control over the instrumental line wing contribution. We find the continuum radiance to range between 60--170\,\phs{}, or 21.4--22.8\,\magarcsec{} depending on the band pass and the time of observation. The zenith scaled airglow continuum radiance is two to four times brighter than the ZL towards the ecliptic poles: the darkest foreground available for ground- and current space-based observatories. The observations carried out under photometric observing conditions establishing a solid measurement for the darkest sky conditions. We observe the continuum to be stable with a modest decay towards the end of night. The presented observations have been sporadically spread around the year, and we are unable to study seasonal variability reported in literature. We promote a long-term observing campaign for its analysis. Our work demonstrates that the VIS-NIR sky continuum can be very dark under photometric conditions, offering close to zodiacal light limited foreground from the ground in the NIR sensitivity range of silicon photo-detectors, encouraging design of astronomical spectrographs taking advantage of the dark interline sky.

\section{\label{sec:availability} Data and code availability}
Raw data is available on the Nordic Optical Telescope's fits archive, or from the corresponding author upon reasonable request. Scripts used as a part of the work are available upon reasonable request from the corresponding author.

\begin{acknowledgements}
The Cosmic Dawn Center (DAWN) is funded by the Danish National Research Foundation under grant DNRF140. The data presented here were obtained in part with ALFOSC, which is provided by the Instituto de Astrofisica de Andalucia (IAA) under a joint agreement with the University of Copenhagen and Nordic Optical Telescope. The COBE datasets were developed by NASA's Goddard Space Flight Center under the guidance of the COBE Science Working Group and were provided by the NSSDC. DSCOVR and GOES data courtesy to National Oceanic (NOAO) and Atmospheric Administration Space Weather Prediction Center (SWPC). $F_{10.7}$ data courtesy to the Dominion Radio Astrophysical Observatory (DRAO). We thank J. Munday, V. Pinter, A. E. T. Viitanen, and A. N. Sørensen for discussion and their valuable feedback on the manuscript.
\end{acknowledgements}

%
%
\bibliographystyle{aa}
\bibliography{aa53726-25}

\begin{appendix}

\section{\label{app:apparent}Observed apparent continuum spectra}

\begin{figure}[h!]
    \begin{center}
    \begin{tabular}{c}
        \includegraphics[width=\hsize]{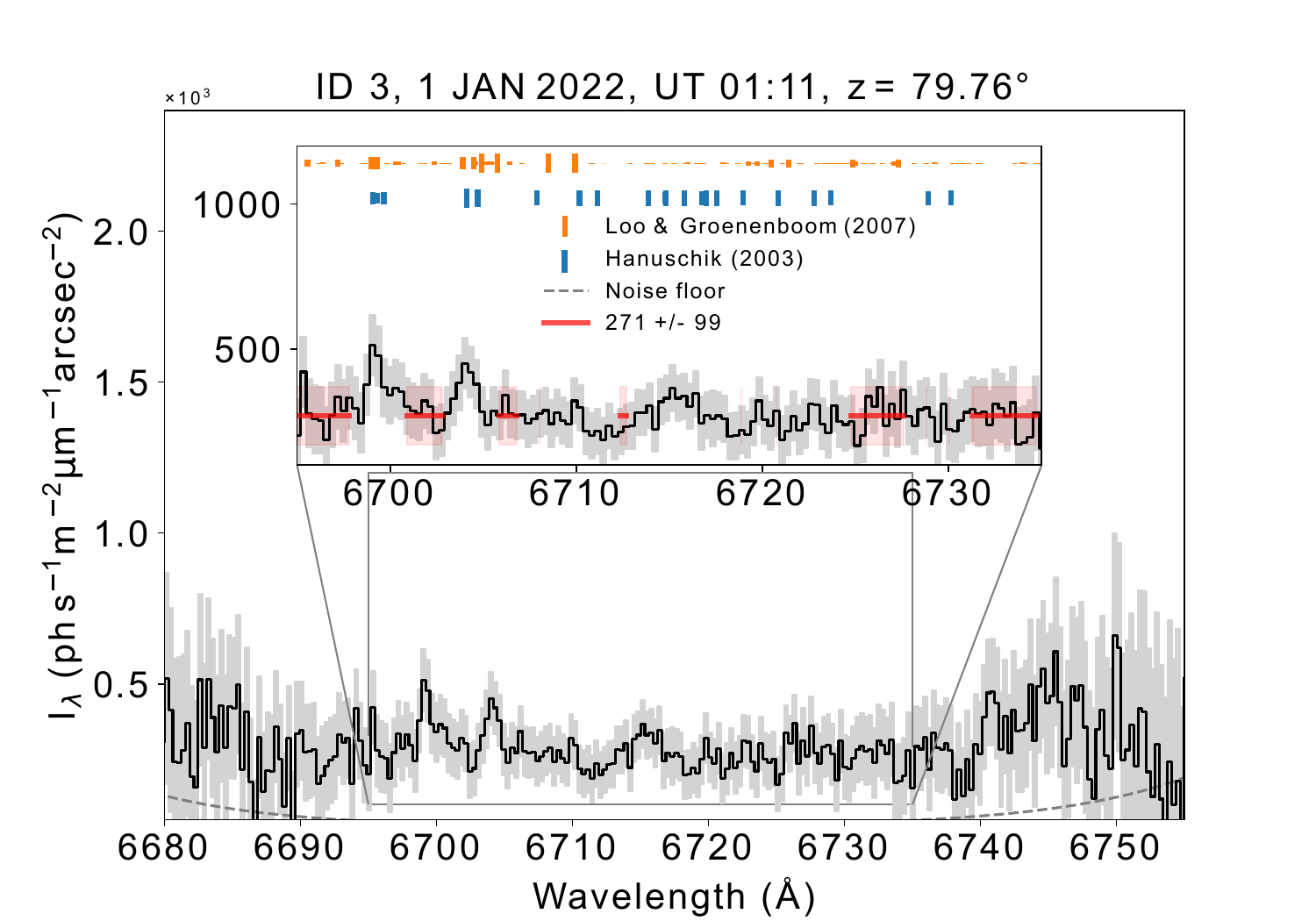} \\
        \includegraphics[width=\hsize]{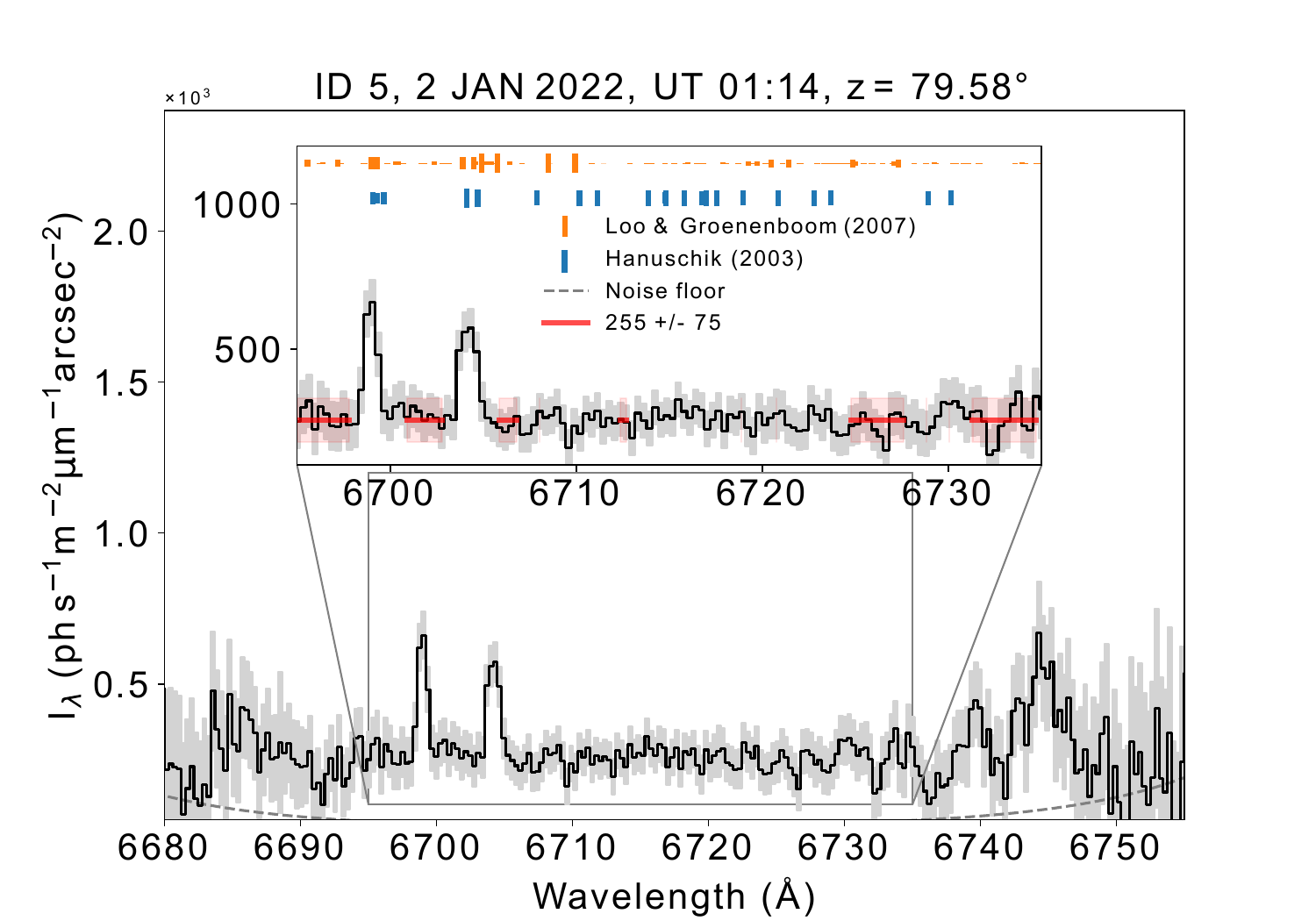} \\
        \includegraphics[width=\hsize]{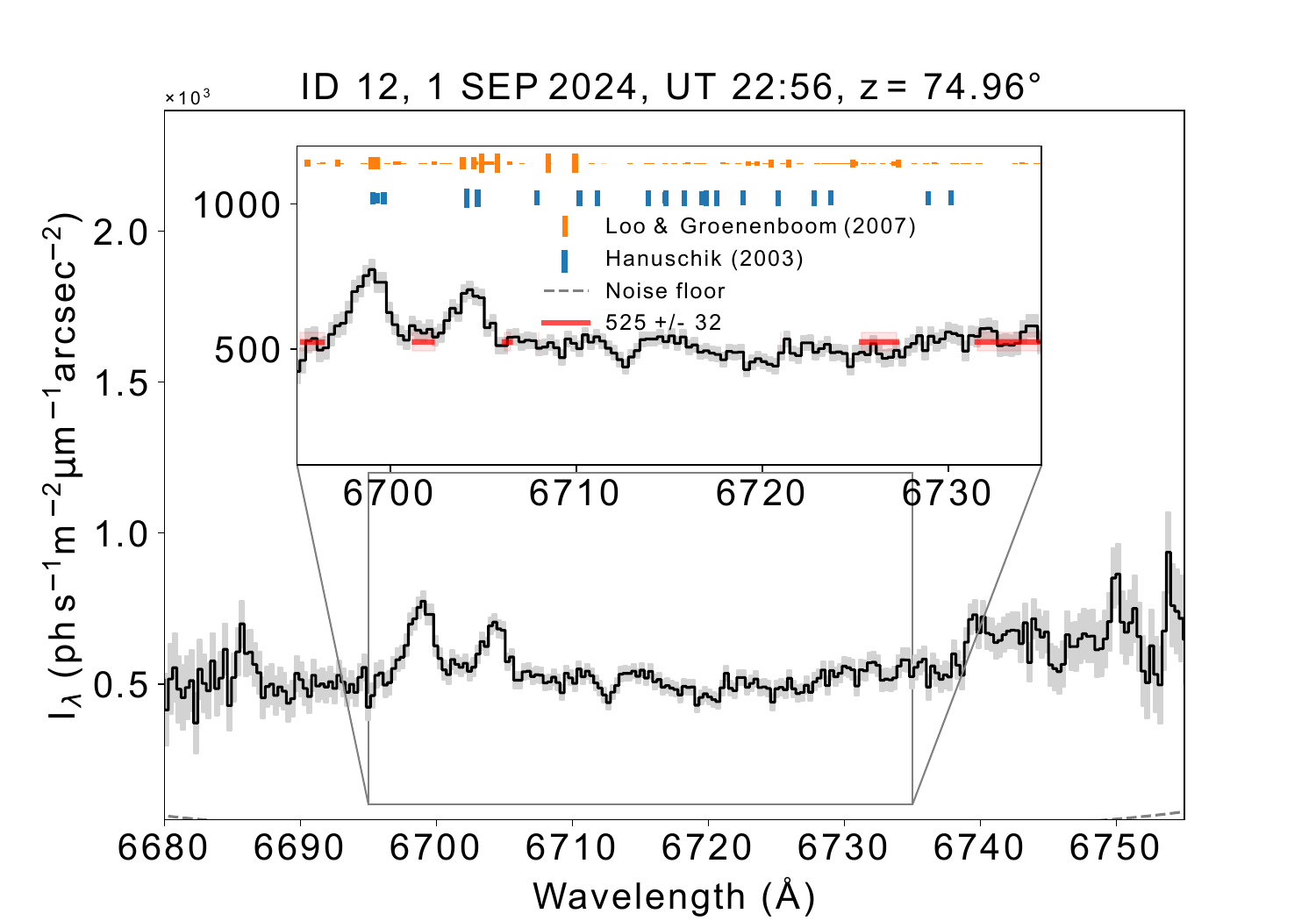} 
    \end{tabular}
    \end{center}
    \caption{\label{fig:app1}Apparent airglow spectra observed in 6720\,{\AA} band. The 2022 spectra are were recorded with 0.5\arcsec slit whereas, the 2024 spectra are with 1.3\arcsec slit. Spectral lines found in \cite{Hanuschik03} are indicated with blue ticks with the tick length indicating relative intensity. Locations of computed OH lines by \cite{Loo07,Loo07erratum} are shown with orange ticks with the tick length indicating the line's Einstein coefficient. The noise floor is shown as a dashed grey line, and spectral regions used for sampling the continuum are indicated with red.
    }
\end{figure} 

\begin{figure}[h!]
    \begin{center}
    \begin{tabular}{c}
        \makebox[\linewidth]{\rule{0pt}{9.5mm}} \\
        \includegraphics[width=\hsize]{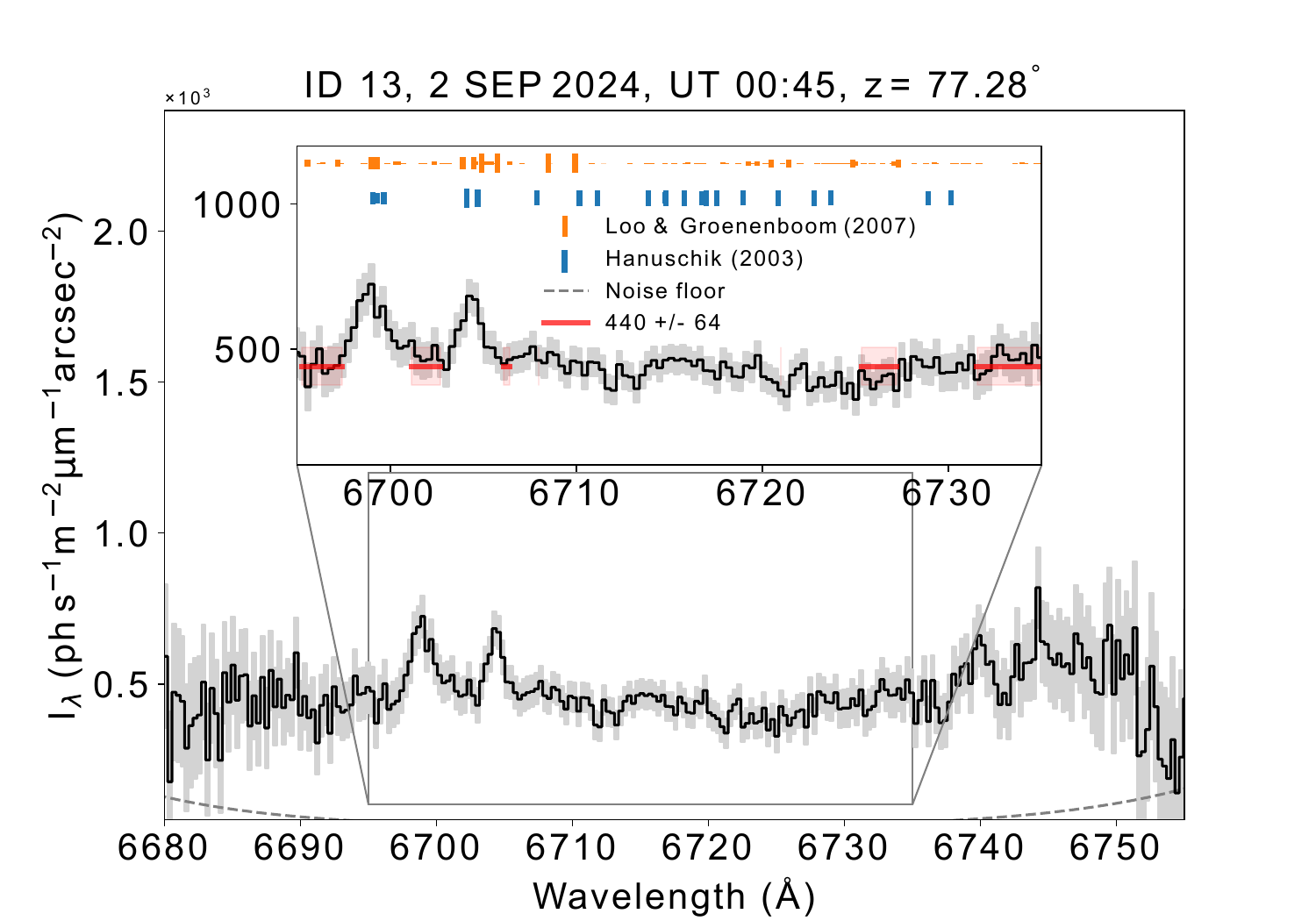} \\
        \includegraphics[width=\hsize]{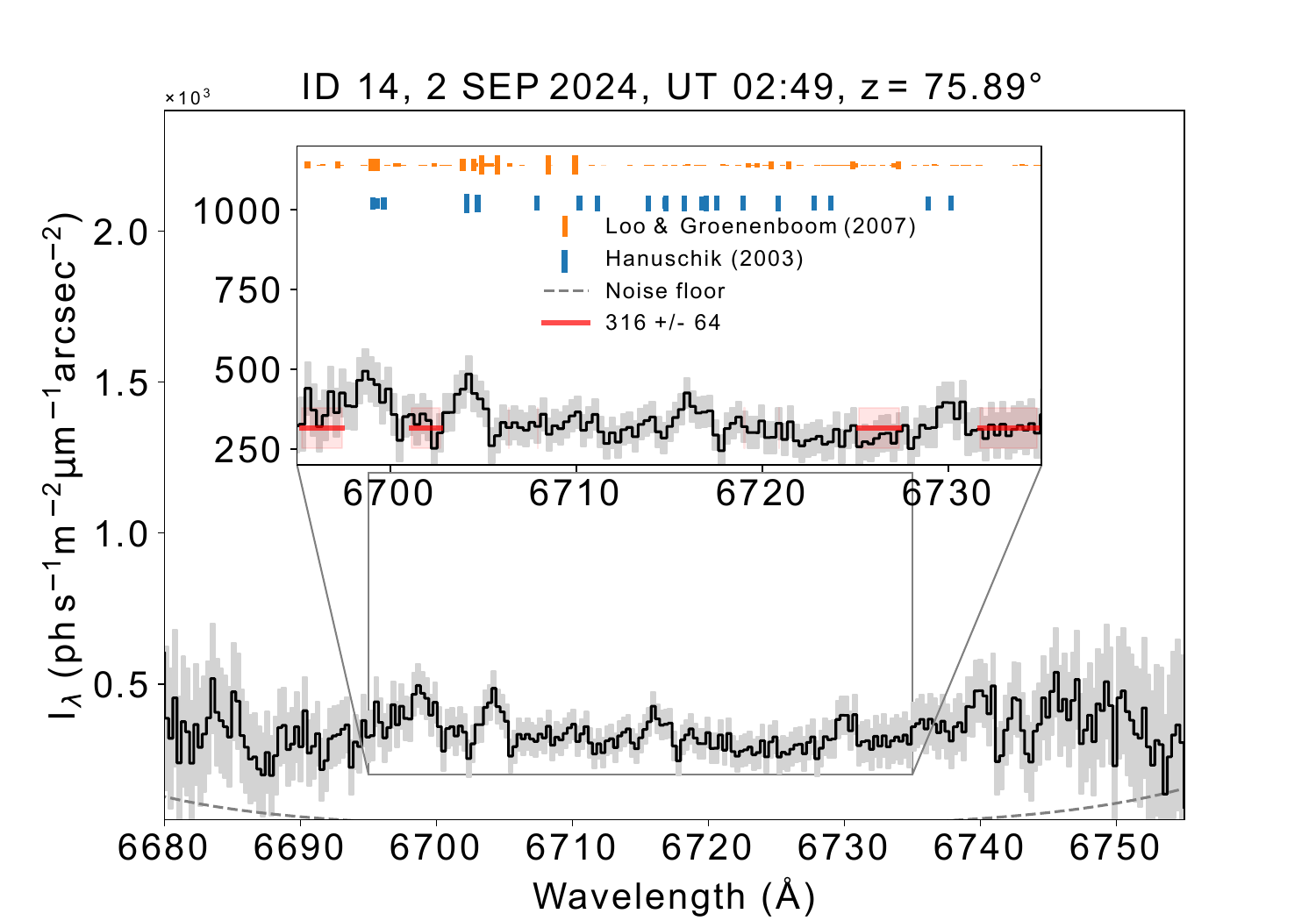} \\
        \includegraphics[width=\hsize]{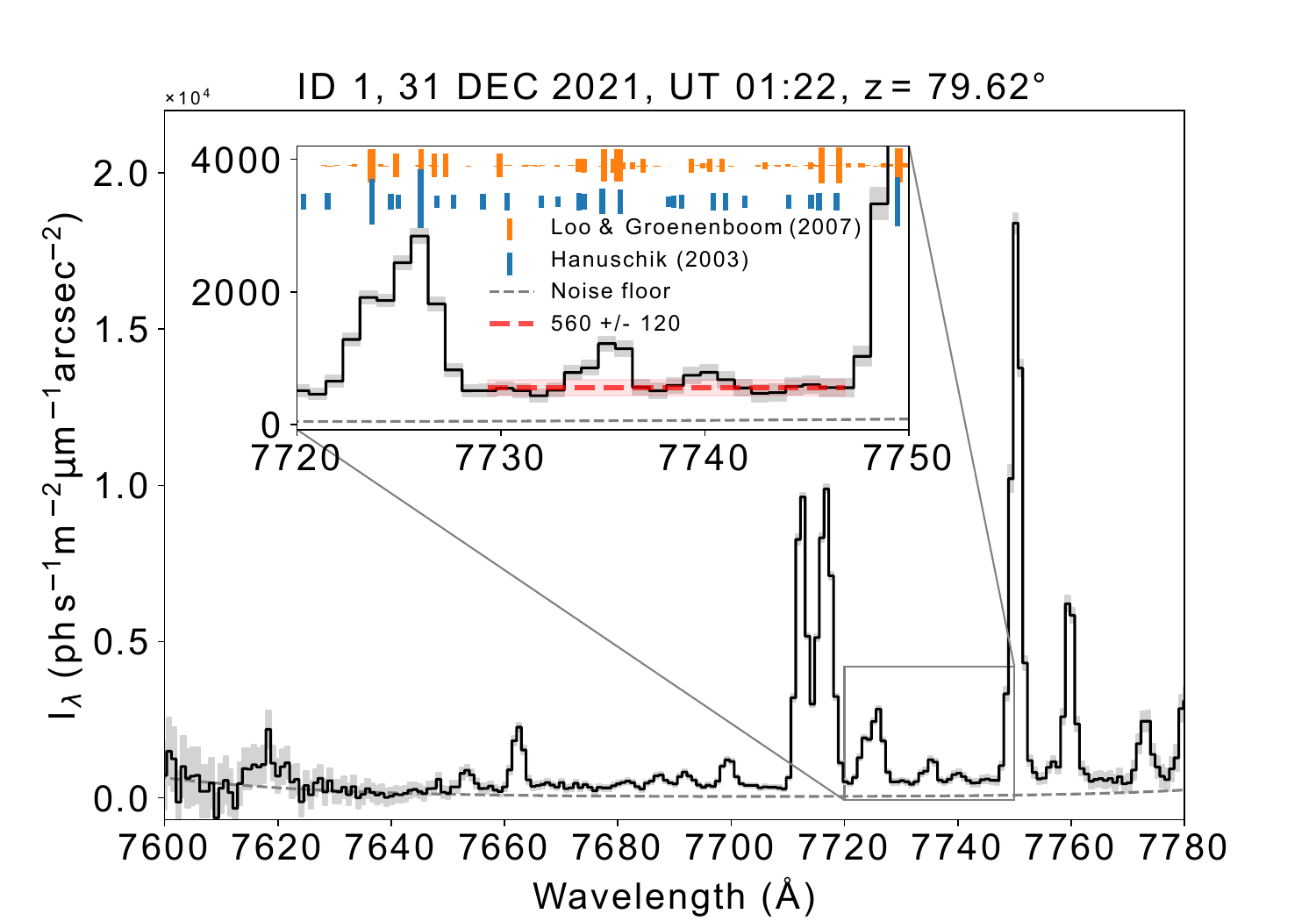}
    \end{tabular}
    \end{center}
    \caption{\label{fig:app2}6720\,{\AA} band spectra continued, and apparent airglow radiance in 7700\,{\AA} band. No clean spectral region is found without known lines, or atmospheric absorption features. An upper limit on the continuum radiance is derived from regions free from \mbox{O$_2$} absorption, but containing unresolved spectral lines.
    }
\end{figure} 

\newpage

\begin{figure}[h!]
    \begin{center}
    \begin{tabular}{c}
        \includegraphics[width=\hsize]{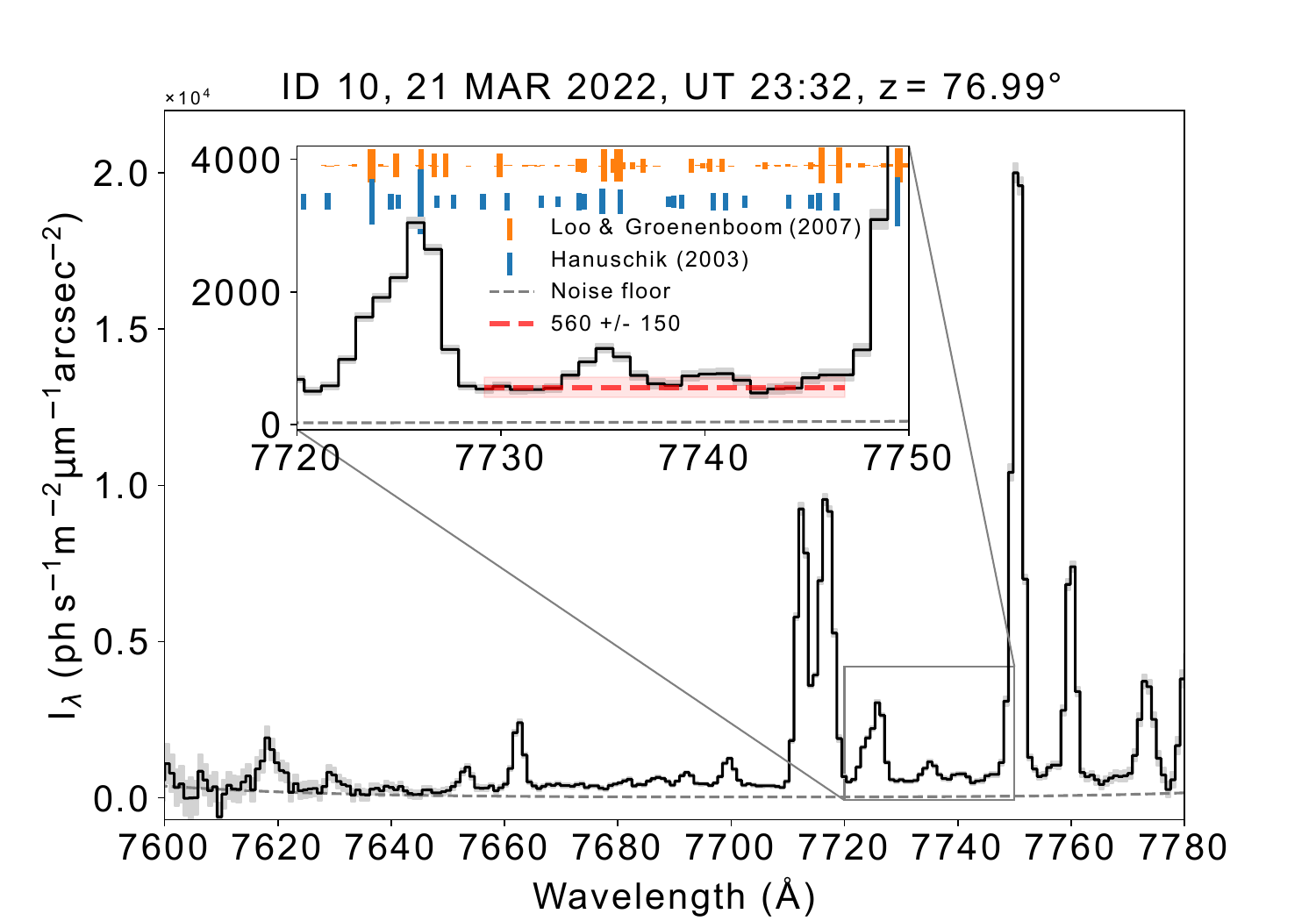} \\
        \includegraphics[width=\hsize]{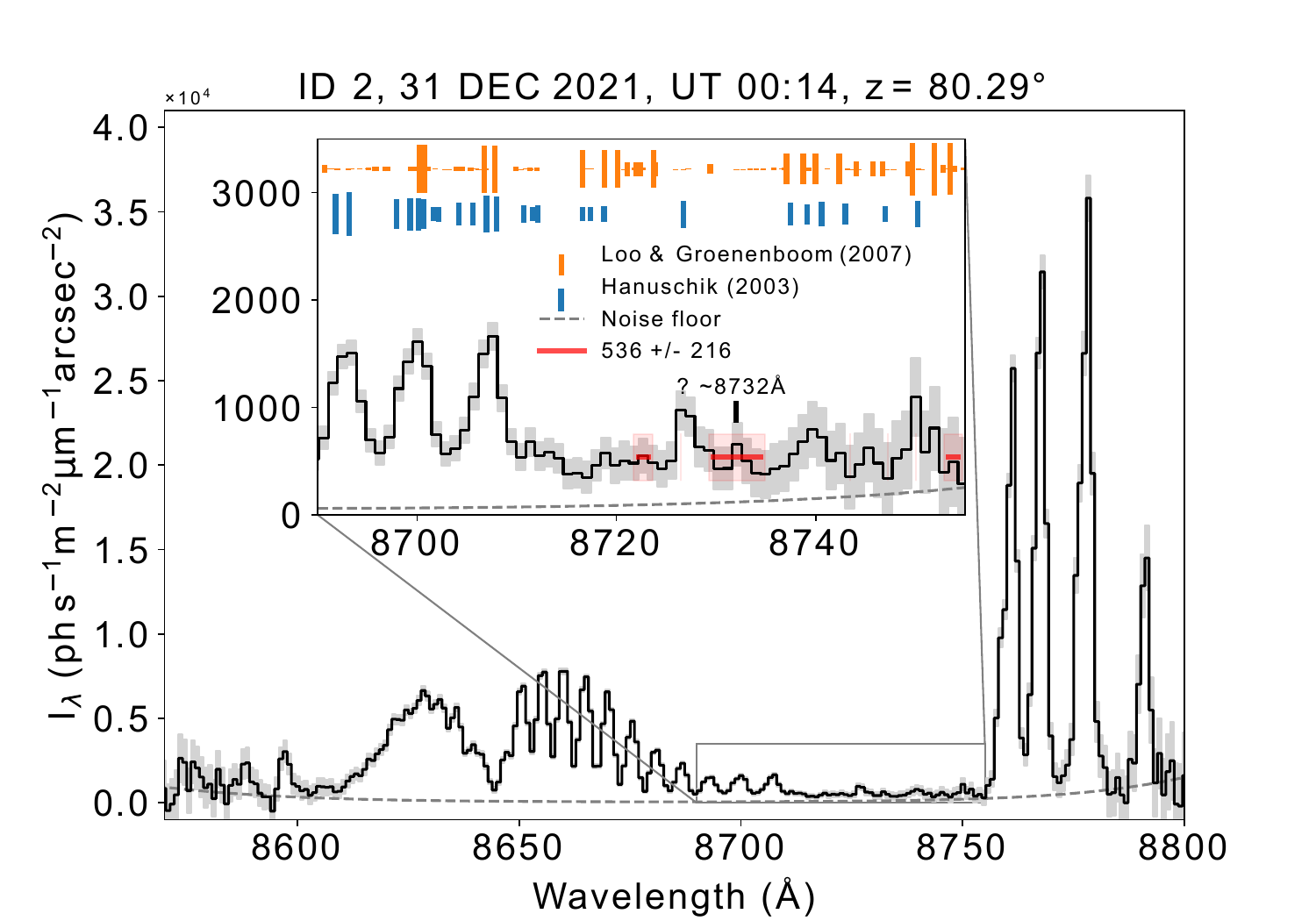}\\
        \includegraphics[width=\hsize]{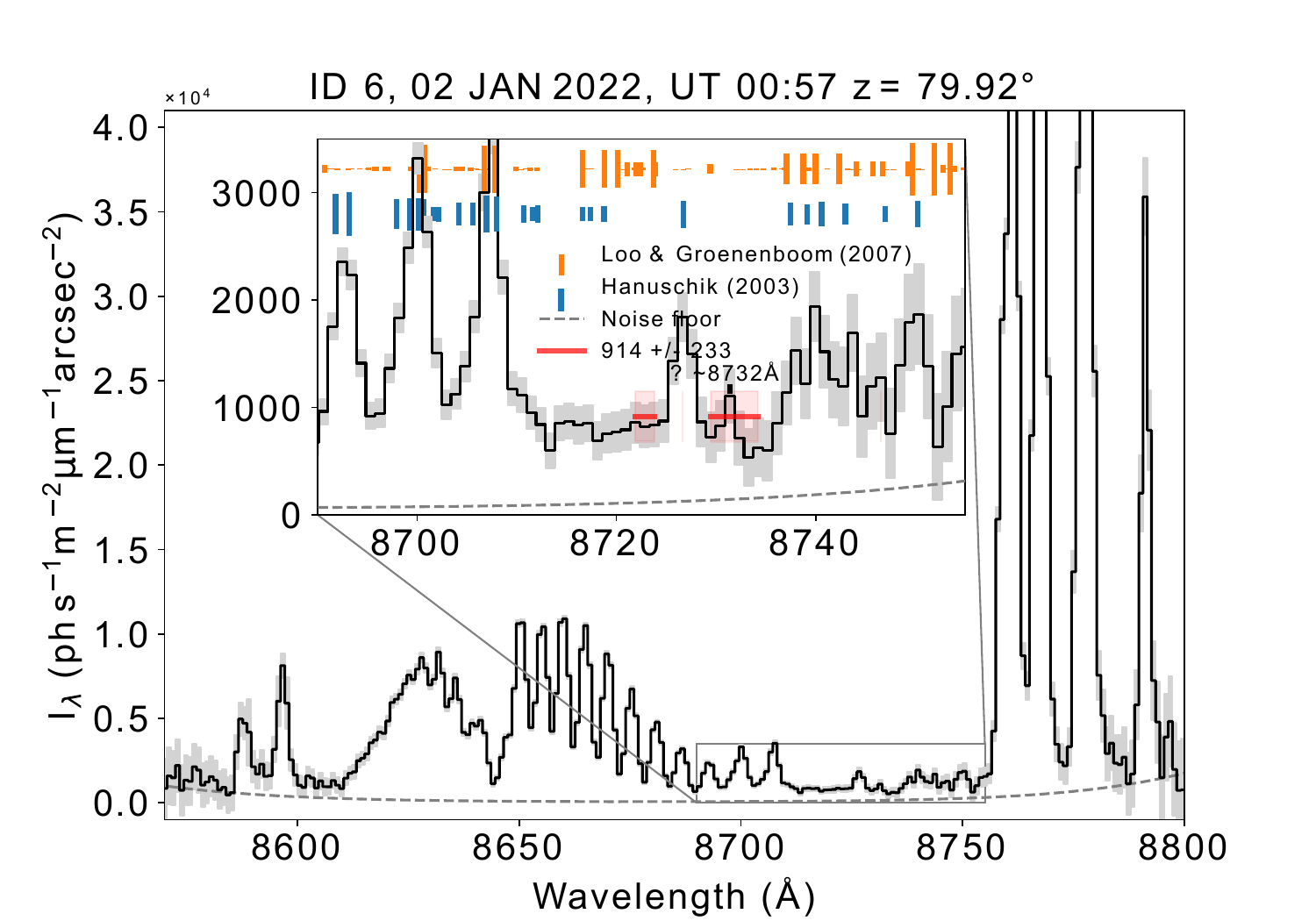}
    \end{tabular}
    \end{center}
    \caption{\label{fig:app3}7700\,{\AA} band airglow spectra continued. Apparent airglow spectra in 8700{\AA} band.
    }
\end{figure} 

\begin{figure}[h!]
    \begin{center}
    \begin{tabular}{c}
        \includegraphics[width=\hsize]{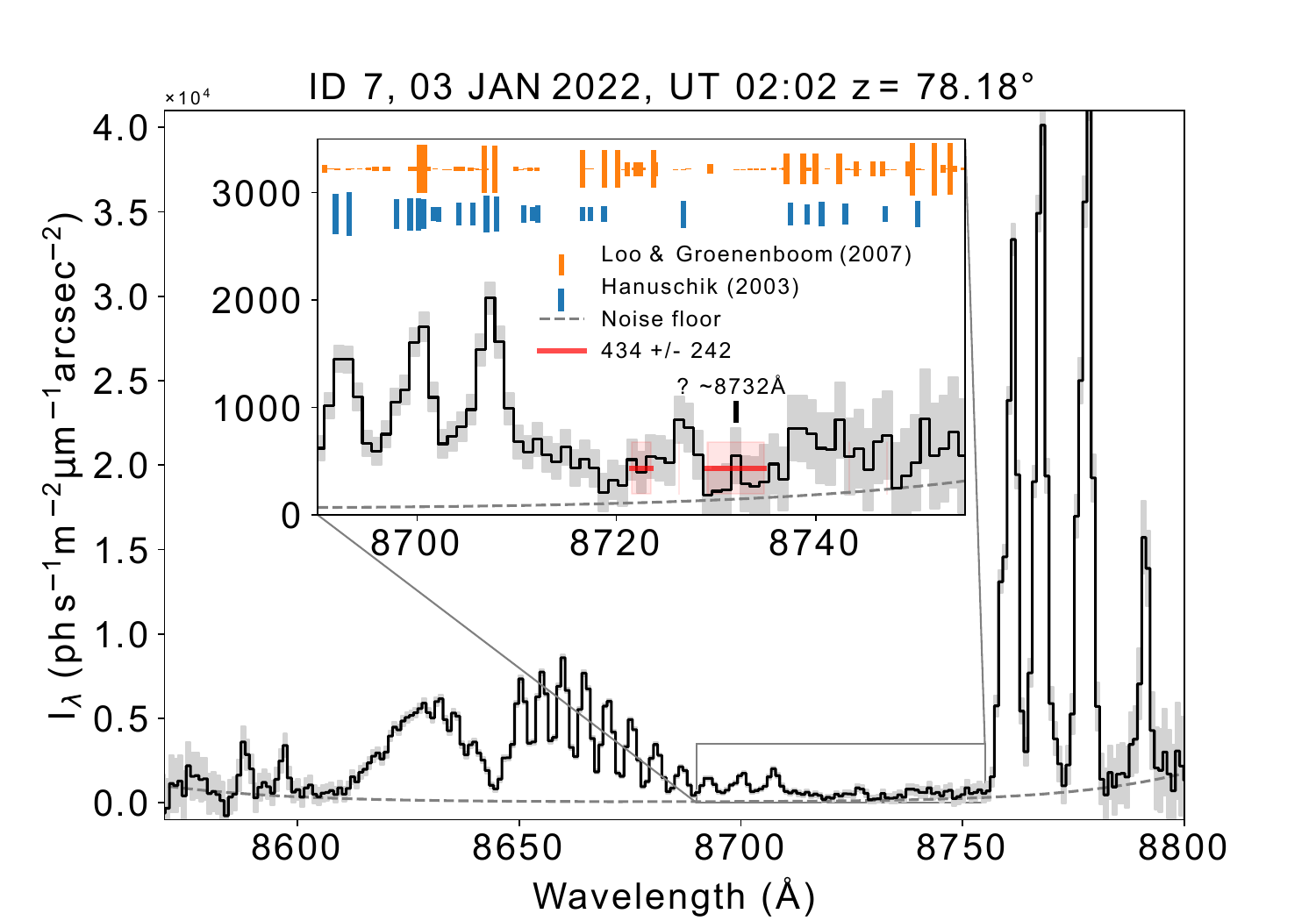} \\
        \includegraphics[width=\hsize]{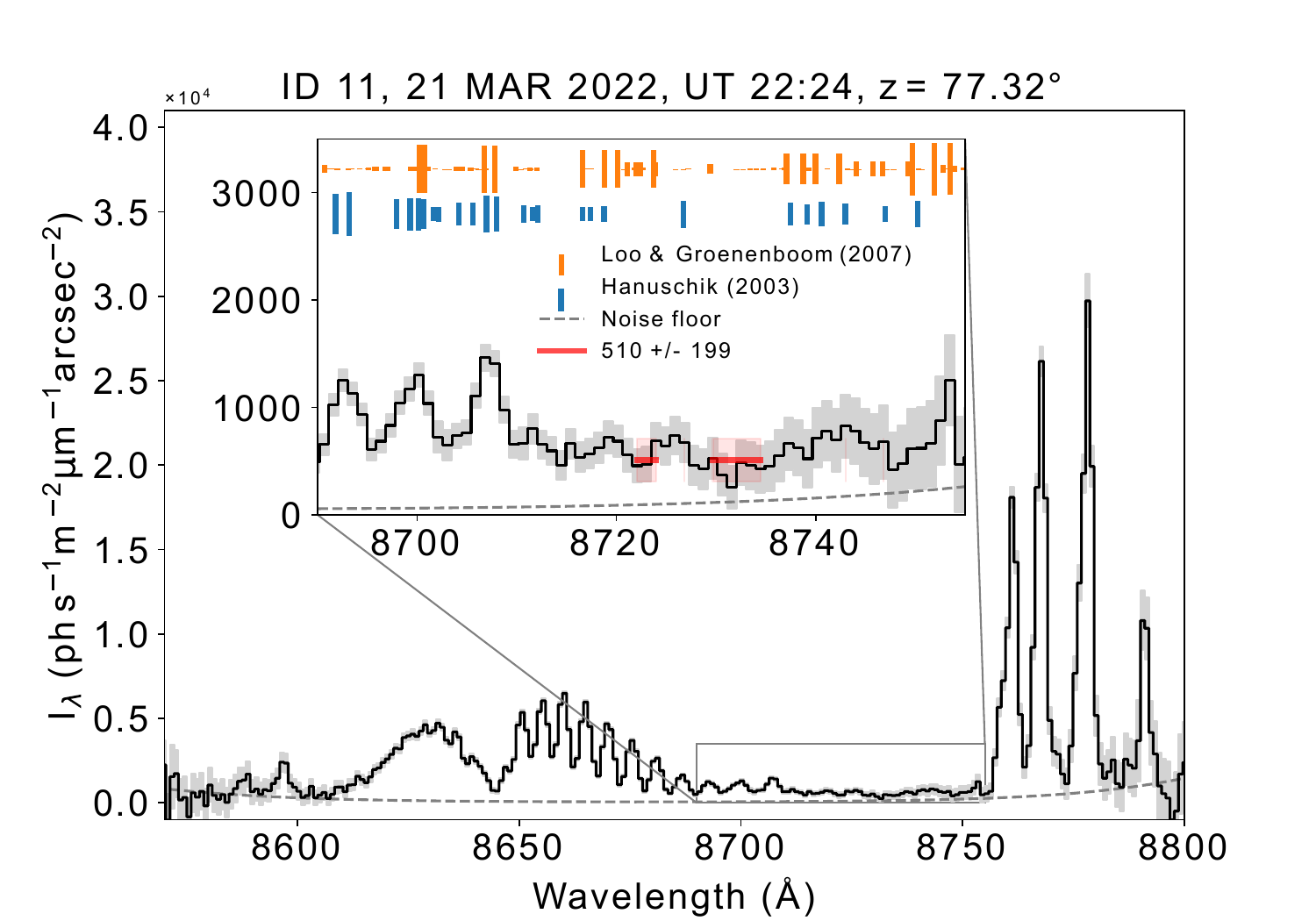} \\
        \includegraphics[width=\hsize]{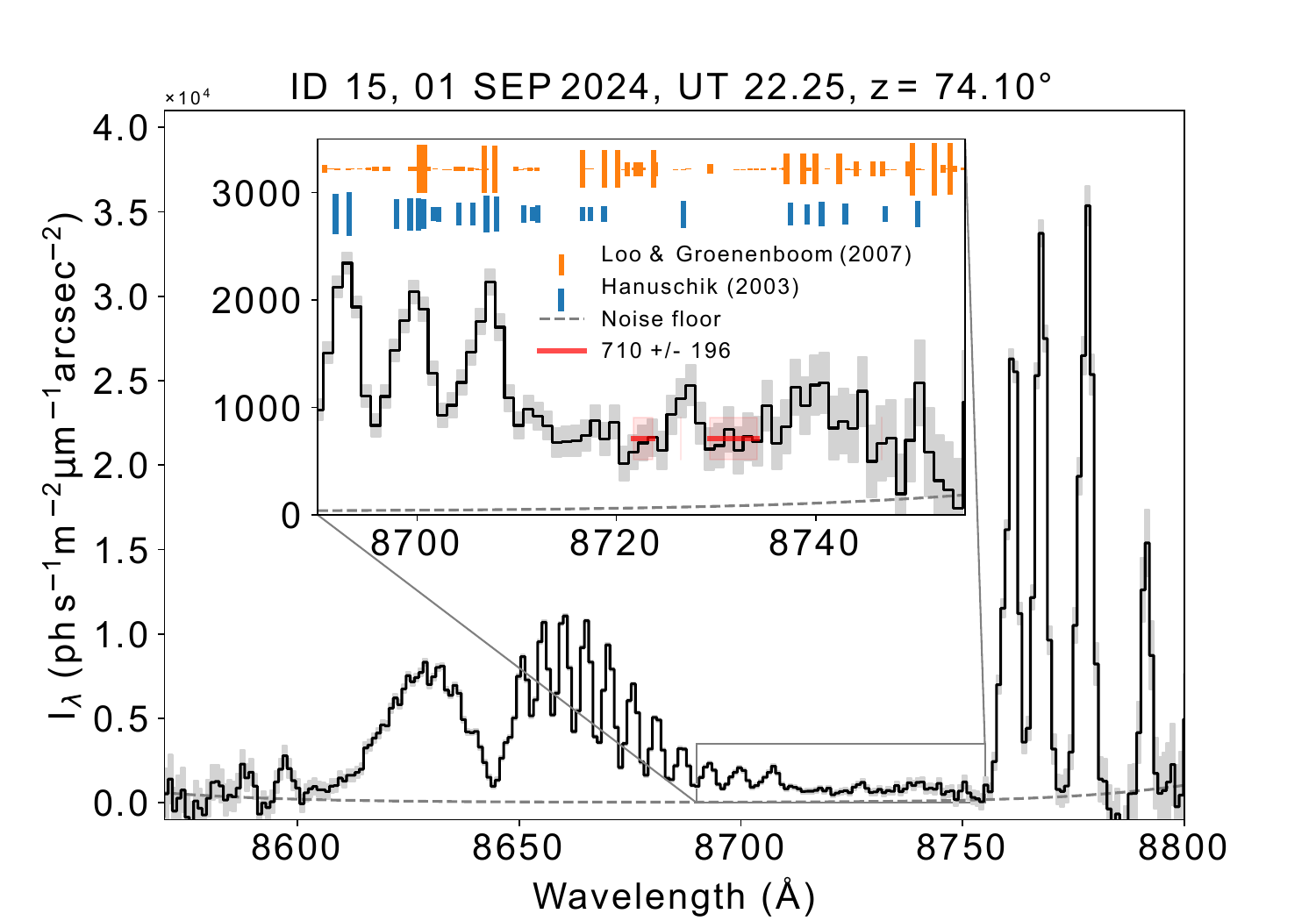} 
    \end{tabular}
    \end{center}
    \caption{\label{fig:app4}Apparent airglow spectra in 8700{\AA} band continued.
    }
\end{figure}

\newpage

\begin{figure}[h!]
    \begin{center}
    \begin{tabular}{c}
        \includegraphics[width=\hsize]{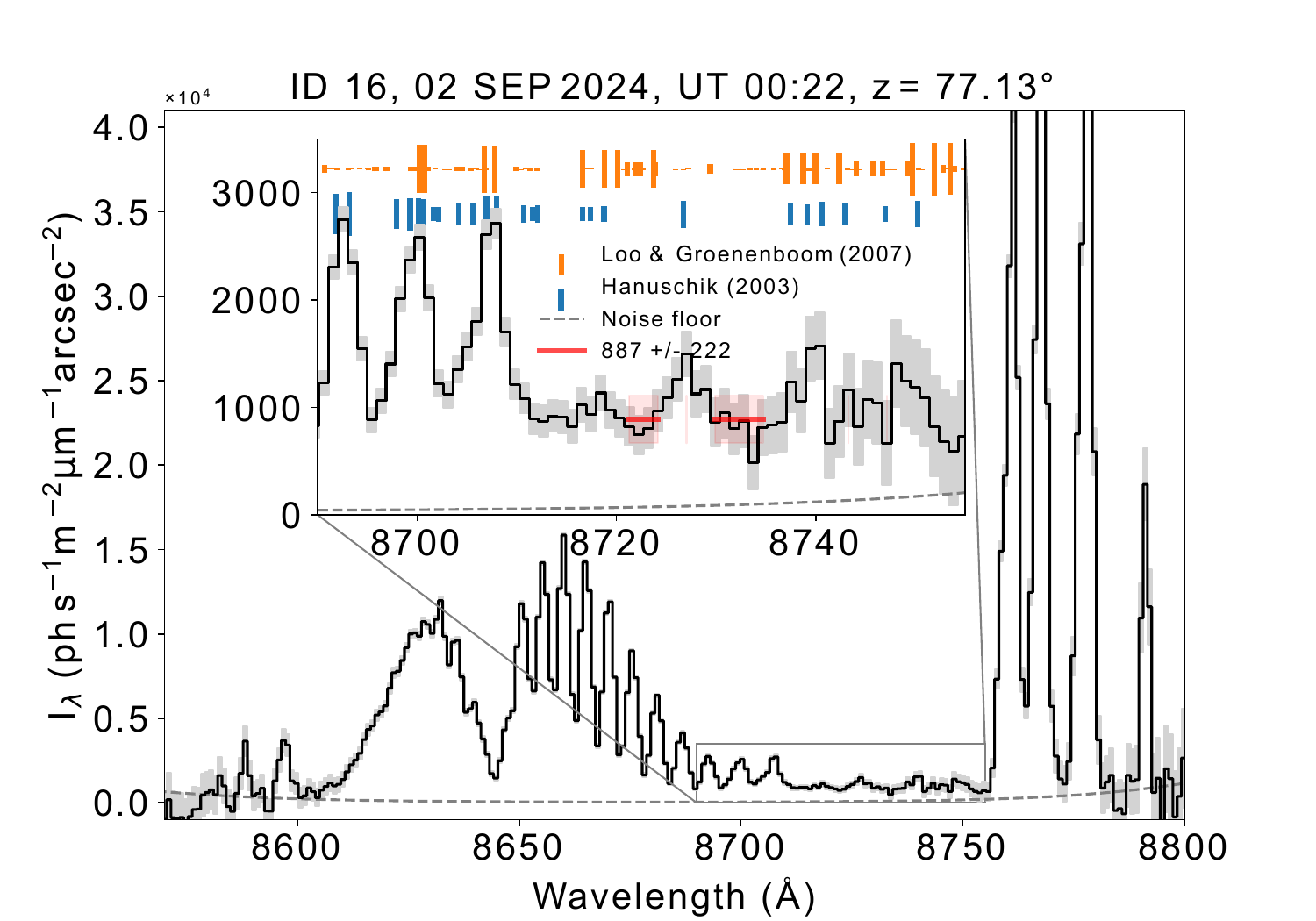} \\
        \includegraphics[width=\hsize]{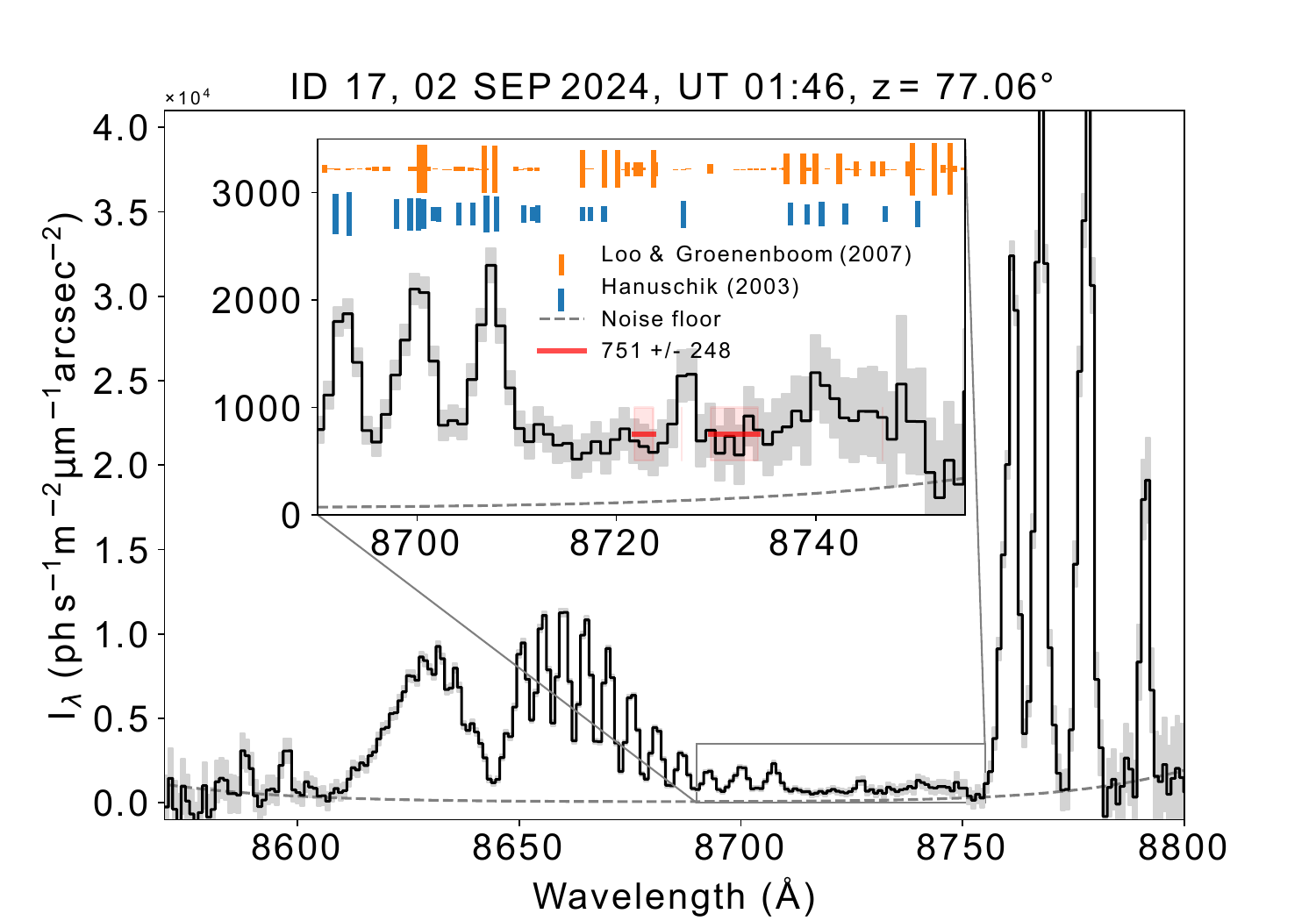} \\
        \includegraphics[width=\hsize]{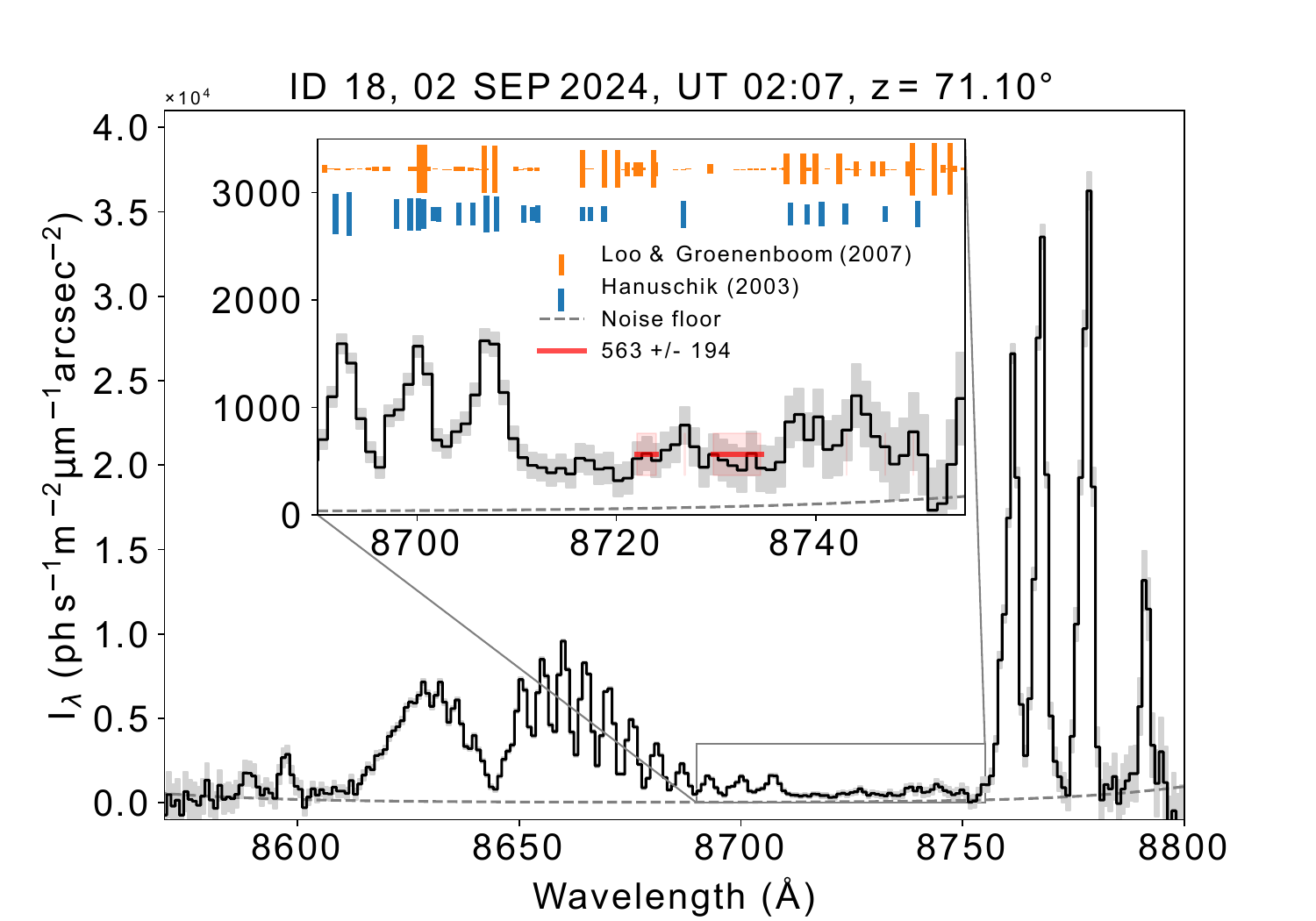} 
    \end{tabular}
    \end{center}
    \caption{\label{fig:app5}Apparent airglow spectra in 8700{\AA} band continued.
    }
\end{figure}

\begin{figure}[h!]
    \begin{center}
    \begin{tabular}{c}
        \includegraphics[width=\hsize]{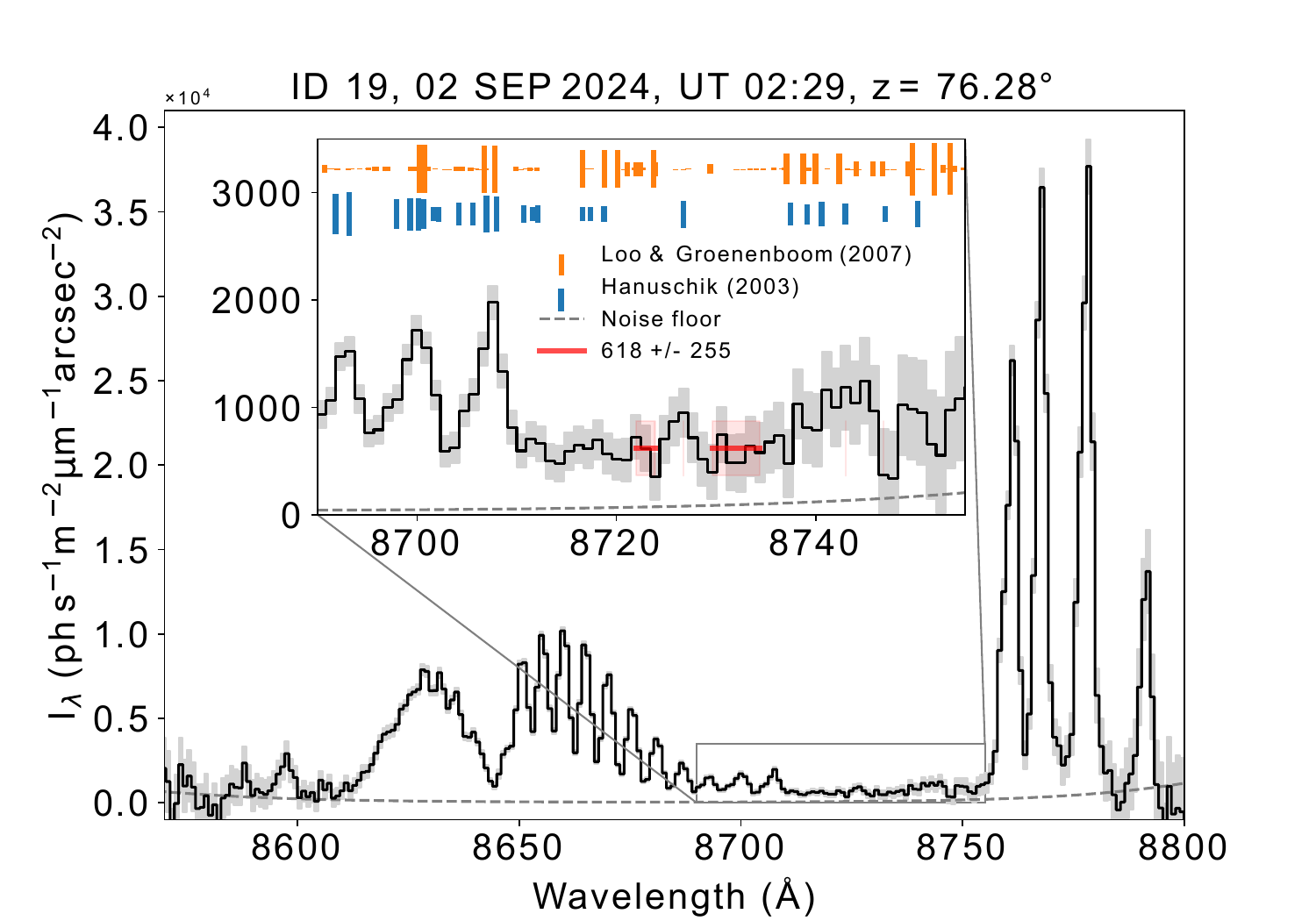}
    \end{tabular}
    \end{center}
    \caption{\label{fig:app6}Apparent airglow spectra in 8700{\AA} band continued.
    }
\end{figure} 

\newpage

\begin{figure}[h!]
    \begin{center}
    \begin{tabular}{c}
        \includegraphics[width=\hsize]{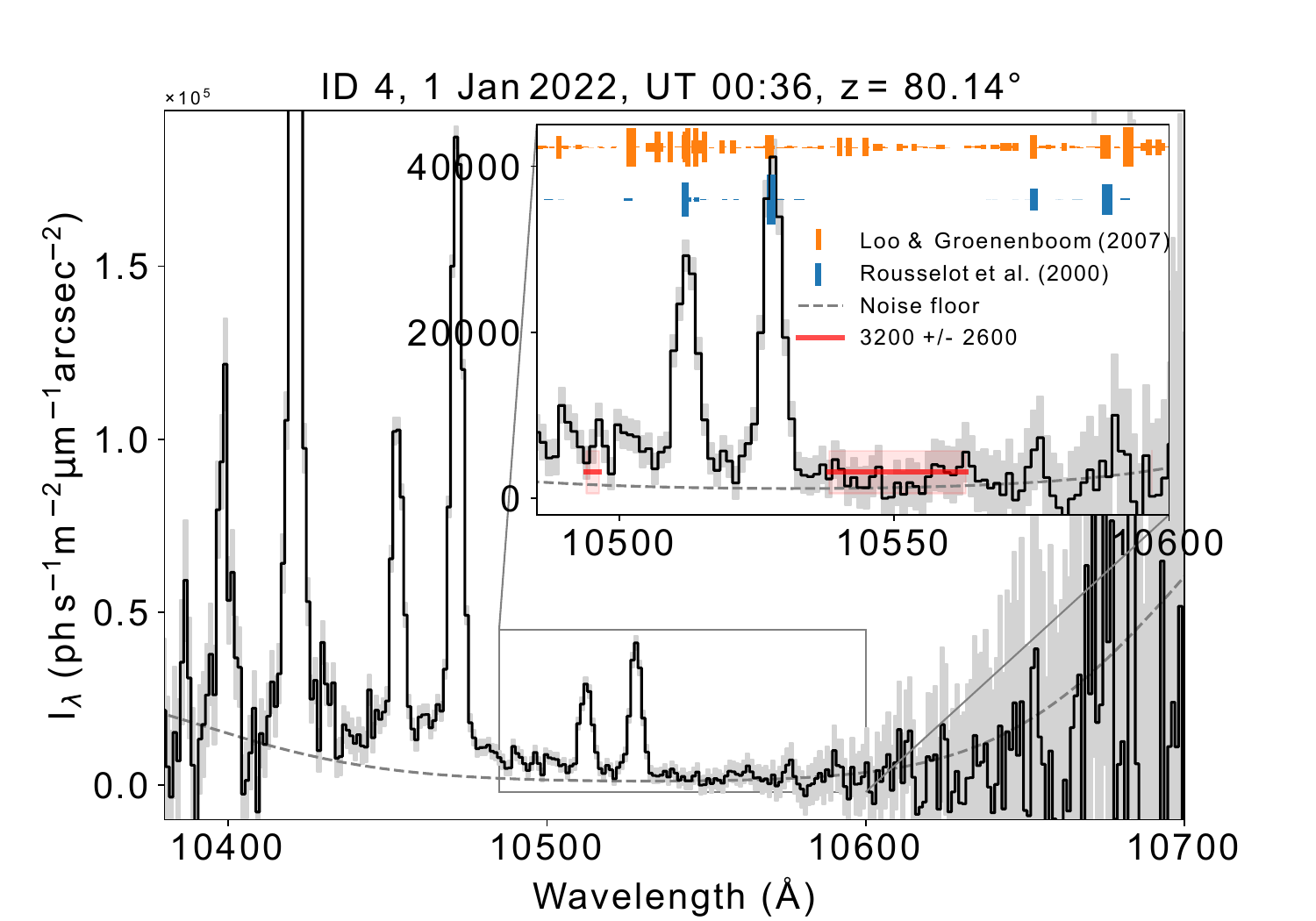} \\
        \includegraphics[width=\hsize]{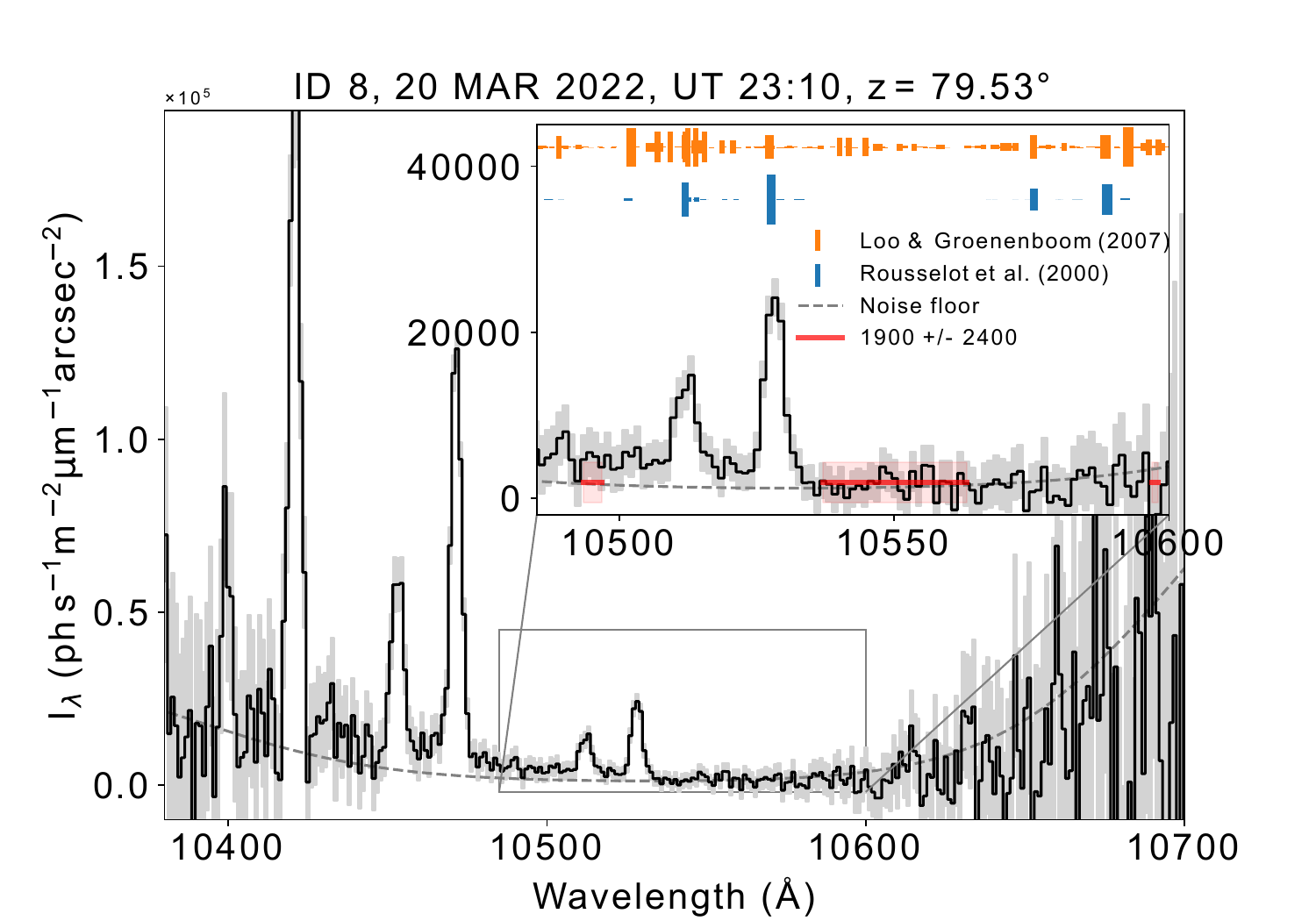} \\
        \includegraphics[width=\hsize]{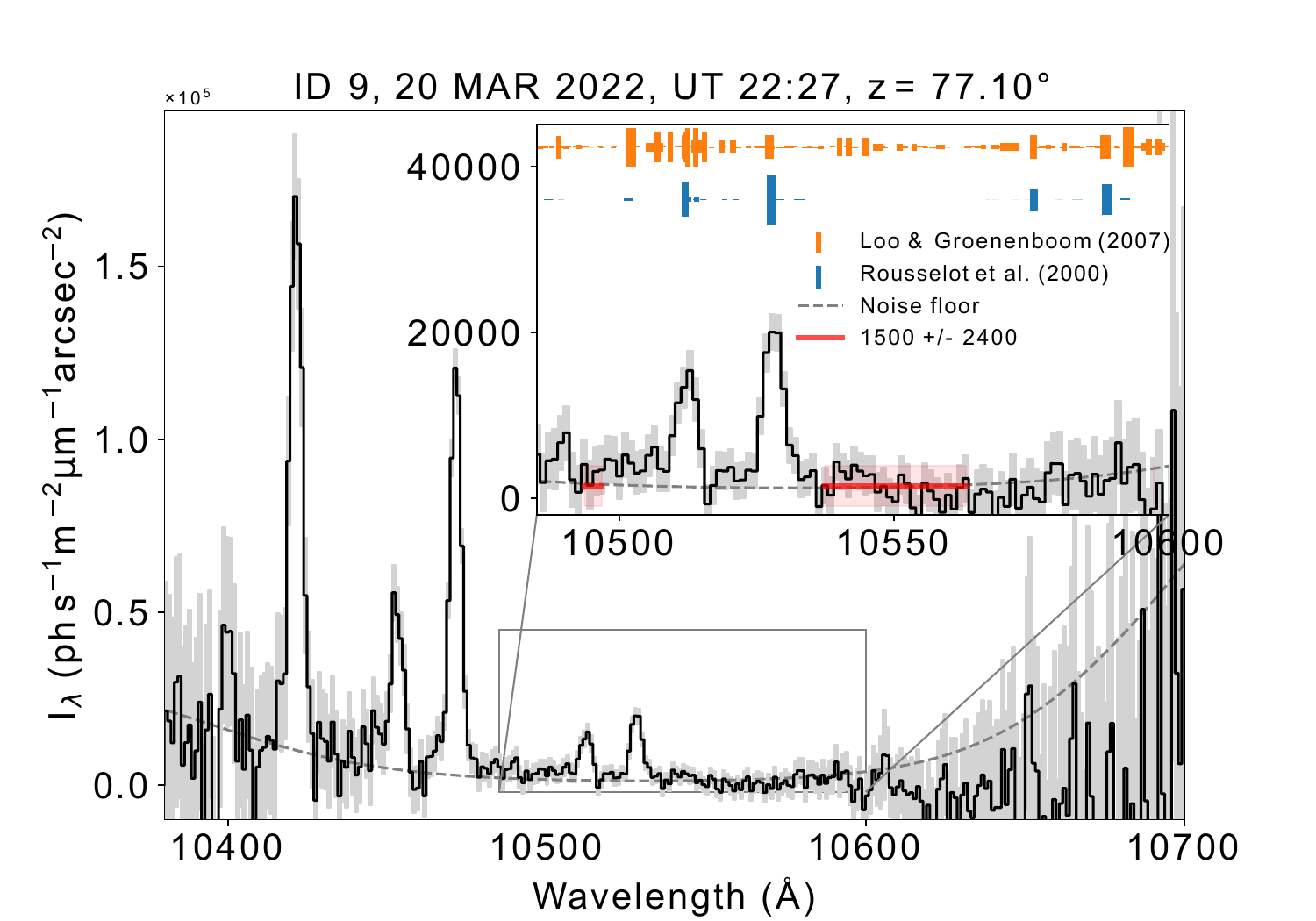}
    \end{tabular}
    \end{center}
    \caption{\label{fig:app7}Apparent airglow spectra in 10\,500\,{\AA} band. Due to low system efficiency, the continuum in 10\,500\,{\AA} is indistinguishable from detector noise in all our spectra.
    }
\end{figure} 


\onecolumn
\section{\label{app:history}Previous studies and solar data}

\begin{table*}[h!]
    \small
    \centering
    \caption{Compilation of NIR and SWIR airglow continuum measurements in the literature.}
    \label{table:previous}
    \begin{tabular}{lrlcccc}
    \hline\hline
                                    &           &             &                      &                      &                          &                      \\
    Publication                     & \multicolumn{2}{c}{Continuum radiance} &  Wavelength / Band & R           & OH Suppression           &\multicolumn{1}{c}{Date} \\
                                    &           &             &                      &                      &                          &                      \\
                                    & \multicolumn{2}{c}{$\mathrm{ph\,s^{-1}m^{-2}\mu m^{-1}}$} & {\AA} / filter& $\lambda/\Delta\lambda$ &                   &                      \\
                                    & \multicolumn{2}{c}{$\mathrm{arcsec^{-2}}$}&    &                      &                          &                      \\
    \hline
                                    &           &             &                      &                      &                          &                      \\
    \cite{Noll24}$^{4\dagger}$      &   375     &             & 16\,550              & 4000 -- 17\,000      & None                     & 2009 -- 2019         \\
                                    &   140     &             & 10\,500              & \ditto{}             & \ditto{}                 & \ditto{}             \\
                                    &   116     &             & 8700                 & \ditto{}             & \ditto{}                 & \ditto{}             \\
                                    &   105     &             & 7700                 & \ditto{}             & \ditto{}                 & \ditto{}             \\
                                    &   80      &             & 6700                 & \ditto{}             & \ditto{}                 & \ditto{}             \\
    \cite{Nguyen16}$^3$             &   751     & $\pm$ 4     & 11\,913              & 320                  & NB filter                & 15 Feb 2013          \\
    \cite{Oliva15}$^1$              &   380     &             & 16\,650              & 32\,000              & None                     & 3 Sep 2014           \\
                                    &   300     &             & H                    & \ditto{}             & \ditto{}                 & \ditto{}             \\
    \cite{Trinh13}$^1$              &   560     & $\pm$ 120   & 15\,200              & 2400                 & FBG                      & 1--5 Sep 2011        \\
    \cite{Ellis12}$^1$              &   860     & $\pm$ 210   & H                    & 2400                 & FBG                      & 1--5 Sep 2011        \\
    \cite{Sullivan12}$^1$           &   670     & $\pm$ 200   & 16\,650              & 6000                 & None                     & Mar 2010             \\
                                    & 1017      & $\pm$ 19    & H                    & \ditto{}             & \ditto{}                 & \ditto{}             \\
                                    &   663     & $\pm$ 19    & J                    & \ditto{}             & \ditto{}                 & \ditto{}             \\
                                    &   508     & $\pm$ 19    & Y                    & \ditto{}             & \ditto{}                 & \ditto{}             \\
    \cite{Tilvi10}                  &   162     &             & 10\,630              & 1300                 & NB filter                & 1--6 Oct 2008        \\
    \cite{Venemans09}               &   750     &             & 10\,600              & 1000                 & NB filter                & 2--10 Nov 2006       \\
    \cite{Hanuschik03}$^1$          &   335     &             & 8600-10\,430         & 45\,000              & None                     & 20--22 June 2001     \\
                                    &   307     &             & 6720-8560            & \ditto{}             & \ditto{}                 & \ditto{}             \\
    \cite{Cuby00}$^1$               & 2300      &             & 11\,700              & 3000                 & None                     & Jan 2000             \\
                                    & 1200      &             & 11\,900              & 3000                 & None                     & \ditto{}             \\
    \cite{Maihara93b}$^1$           &   590     & $\pm$ 140   & 16\,650              & 1900                 & Mirror mask              & Feb 1992             \\
    \cite{Sobolev78}$^2$            &   296     & $\pm$ 64    & 10\,612              & 12{\AA}-mm, 9deg fov & \ditto{}                 & late 1976            \\
                                    &   144     & $\pm$ 57    & 9268                 & \ditto{}             & \ditto{}                 & \ditto{}             \\
                                    &   280     & $\pm$ 94    & 8210                 & \ditto{}             & \ditto{}                 & \ditto{}             \\
    \cite{Noxon78}$^2$              &   131     & $\pm$ 94/19 & 8570                 & N/A                  & None                     & 15 Sep 1977          \\
    \cite{Gadsden73}$^{4\dagger}$   &    56     &             & 7154                 & 160                  & NB filter                & \ditto{}             \\
                                    &    34     &             & 6754                 & 240                  & \ditto{}                 & \ditto{}             \\
    \cite{Sternberg72}$^{4\dagger}$ &   280     & $\pm$ 94    & 8200                 & N/A                  & None                     & 18 Aug 1969          \\
    \cite{Broadfoot68}$^{1\dagger}$ &  990      & $\pm$ 230   & 8700                 & $\sim$5000?          & None                     & N/A                  \\
                                    &  200      & $\pm$ 70    & 7700                 & \ditto{}             & \ditto{}                 & \ditto{}             \\
                                    &  103      & $\pm$ 15    & 6700                 & \ditto{}             & \ditto{}                 & \ditto{}             \\
    \cite{Krassovsky62}             &  374      &             & 4000-7000            & N/A                  & None                     & N/A                  \\

    \end{tabular}
    \tablefoot
    {\tiny Additional works exist at bluer wavelengths. Results are reported in various units in the original works and are converted to match the units used in this work. Some of the values have been read from figures, few of which have been photocopied. \tablefoottext{$\dagger$}{Authors report broader wavelenght coverage, table values have been chosen to match the wavelengths studied in this work.}\tablefoottext{1}{Apparent,} \tablefoottext{2}{scaled to zenith}, \tablefoottext{3}{ZL subtracted but not zenith scaled,} \tablefoottext{4}{ZL subtracted and zenith scaled.}}
\end{table*}

\begin{table*}[h!]
\small
\centering
\caption{Space weather conditions at the time of observation.}
\label{table:solardata}

\begin{tabular}{ccccccc}
\hline\hline
    &               &                &               &                             &                                  &            \\
Id  & Solar wind    & Solar wind     & Mag. field    & Solar X-ray                 & >10MeV                           & DRAO       \\
    & speed         & density        & Bz            & flux                        & protons                          & $F_{10.7}$ \\
    &               &                &               &                             &                                  &            \\
    & \mbox{km\,s$^{-1}$} & \mbox{cm$^{-3}$} & \mbox{nT} & \wsqm{}                 & $\mathrm{p^{+}\,s^{-1}\,cm^{-2}\,sr^{-1}\,keV^{-1}}$                     & \mbox{sfu} \\
\hline
    &               &                &                &                            &                                 &            \\
2   & 465$\pm$2     & 8.5$\pm$0.2    & ~2.2$\pm$0.9   & 2.69$\pm0.02\times10^{-7}$ & ~\,3.8$\pm2.6\times10^{-5}$     & 103.2      \\
6   & ~~542$\pm$25  & 11.5$\pm$2.0   & -2.7$\pm$3.8   & 4.53$\pm0.46\times10^{-7}$ & ~\,4.6$\pm3.5\times10^{-5}$     & 93.3       \\
7   & 524$\pm$4     & 13.1$\pm$1.0   & ~2.5$\pm$1.8   & 2.25$\pm0.32\times10^{-7}$ & ~\,4.3$\pm2.9\times10^{-5}$     & 88.3       \\
11  & 522$\pm$8     & 10.2$\pm$1.0   & -0.9$\pm$2.5   & 8.17$\pm0.41\times10^{-7}$ &   13.6$\pm6.9\times10^{-5}$     & 96.2       \\
15  & 427$\pm$5     & 6.2$\pm$0.3    & ~2.1$\pm$2.4   & 5.11$\pm0.65\times10^{-6}$ & ~\,5.2$\pm2.5\times10^{-5}$     & 231.1      \\
16  & 432$\pm$8     & 5.9$\pm$0.7    & -0.7$\pm$0.8   & 4.75$\pm0.14\times10^{-6}$ & ~\,9.5$\pm3.6\times10^{-5}$     & 231.1      \\
17  & 446$\pm$3     & 7.5$\pm$0.4    & ~2.8$\pm$1.1   & 4.02$\pm0.05\times10^{-6}$ &   11.8$\pm4.6\times10^{-5}$     & 231.1      \\
18  & 451$\pm$7     & 7.7$\pm$0.5    & ~2.5$\pm$0.9   & 4.13$\pm0.09\times10^{-6}$ &   13.2$\pm4.0\times10^{-5}$     & 231.1      \\
19  & 433$\pm$3     & 6.4$\pm$0.3    & -3.5$\pm$0.6   & 5.06$\pm0.40\times10^{-6}$ &   13.0$\pm5.5\times10^{-5}$     & 231.1      \\
\end{tabular}
\tablefoot
{\tiny Bz in geocentric solar magnetic coordinates. DSCOVR and GOES data courtesy to National Oceanic (NOAO) and Atmospheric Administration Space Weather Prediction Center (SWPC). $F_{10.7}$ data courtesy to Dominion Radio Astrophysical Observatory (DRAO).}
\end{table*}

\end{appendix}
\end{document}